\documentstyle[12pt,a4wide,epsfig,amssymb]{article}

\def\lc{\Lambda_c}
\def\NF{N\hspace{-0.35mm}F}
\def\mk{m_K}
\def\mp{m_\pi}
\def\me{m_\eta}
\def\mep{m_{\eta'}}
\def\nn{\nonumber\\}
\def\Log{\log\Lambda_{c}^2}
\def\directcp{\varepsilon'/\varepsilon}
\def\D{\displaystyle}
\def\T{\textstyle}

\def\SS{\scriptscriptstyle}
\def\lf{\frac{i}{(2\pi)^4}}
\def\Integ{\int \hspace{-1.5mm}d^4q\hspace{1mm}}

\title{
\vspace*{-1.0cm}
\begin{flushright}
{\normalsize DO--TH 98/23\\[-12pt]
LNF-98/044 (P)}
\end{flushright}
\vspace{1.2cm}
{\Large \bf
New Analysis of the \boldmath $\Delta I = 1/2$ \unboldmath Rule in 
Kaon Decays and the \boldmath $\hat{B}_K$ \unboldmath Parameter} 
\vspace*{0.8cm}
}
\author{ 
T. Hambye$^{a\,}$\footnote{\small E-mail: hambye@lnf.infn.it}~, 
G. O. K\"ohler$^{b\,}$\footnote{\small E-mail: 
koehler@doom.physik.uni-dortmund.de}~, and P. H. Soldan$^{b\,}
$\footnote{\small E-mail: soldan@doom.physik.uni-dortmund.de}\\[0.5cm]
\normalsize $a$: {\it INFN - Laboratori Nazionali di Frascati,
P.O. Box 13, I-00044 Frascati, Italy}\\[1mm] 
\normalsize $b$: {\it Institut f\"ur Physik, Universit\"at Dortmund 
D-44221 Dortmund, Germany}\\[2cm] 
}
\date{}
\begin{document}
\maketitle
\thispagestyle{empty}
\vspace*{-1.5cm}
\begin{abstract}
We present a new analysis of the $\Delta I = 1/2$ rule in $K\rightarrow
\pi\pi$ decays and the $\hat{B}_K$ parameter. We use the $1/N_c$ expansion 
within the effective chiral lagrangian for pseudoscalar mesons and compute 
the hadronic matrix elements at leading and next-to-leading order in the 
chiral and the $1/N_c$ expansions. Numerically, our calculation reproduces 
the dominant $\Delta I = 1/2\,$ $K\rightarrow\pi\pi$ amplitude. Our result 
depends only moderately on the choice of the cutoff scale in the chiral loops. 
The $\Delta I = 3/2$ amplitude emerges sufficiently suppressed but shows a 
significant dependence on the cutoff. The $\hat{B}_K$ parameter turns out 
to be smaller than the value previously obtained in the $1/N_c$ approach. 
It also shows a significant dependence on the choice of the cutoff scale. 
Our results indicate that corrections from higher order terms and/or higher 
resonances are large for the $\Delta I = 3/2$ $K\rightarrow\pi\pi$ amplitude 
and the $(|\Delta S|=2)$ $K^0 - \bar{K}^0$ transition amplitude. 
\end{abstract}
\newpage
%
\section{Introduction \label{intro}}
Over the last few decades the kaon system has provided us with a rich field
of phenomenology which has been important for developing our theoretical
understanding of the interplay of weak and strong interactions. The
nonleptonic kaon decays are especially interesting because they provide a
testing ground for QCD dynamics at long distances. Two outstanding
problems in the field are the explanation of the $\Delta I = 1/2$ 
rule in $K\rightarrow\pi\pi$ decays and the calculation of the 
$\hat{B}_K$ parameter which measures the non-perturbative contributions to 
the $(|\Delta S|=2)$ $K^0 - \bar{K}^0$ transition amplitude. An accurate
knowledge of $\hat{B}_K$ is necessary for theoretically investigating the
indirect CP violation in the neutral kaon mass matrix, as well as, the
$K_L-K_S$ mass difference. The $\Delta I=1/2$ rule is particularly
important because it gives rise to the small value of the ratio 
$\directcp$ which measures the direct CP violation in the 
$K\rightarrow \pi\pi$ decay amplitudes.

Since its first observation more than 40 years ago \cite{GPBA} the $\Delta 
I=1/2$ enhancement has attracted a great deal of theoretical interest trying 
to find the dynamical mechanism behind the approximate isospin selection
rule, in particular within the standard model. Experimentally, the 
ratio of the $\Delta I=1/2$ and $\Delta I=3/2$ amplitudes in $K\rightarrow 
\pi\pi$ decays corresponding to $I=0$ and $I=2$ in the final state, 
respectively, was measured to be 
\begin{equation}
\frac{1}{\omega}\,\equiv\,\frac{\mbox{Re}a_0}{\mbox{Re}a_2}\,\equiv\,
\frac{\mbox{Re}(K\rightarrow(\pi\pi)_{I=0})}{\mbox{Re}(K\rightarrow(\pi
\pi)_{I=2})}\,=\,22.2\pm 0.1\,,
\end{equation}
with $A_I=a_I\exp(i\delta_I)$ and $\delta_I$ the final state interaction
phases. This result was particularly enigmatic before the advent of QCD when 
only the current-current operator $Q_2$ arising from the $W$ exchange was
included in the analysis and, consequently, $\mbox{Re}a_0/\mbox{Re}a_2$ 
was expected to be around one. With the establishment of QCD our
understanding of the $\Delta I=1/2$ selection rule improved considerably.
Using the operator product expansion, the $K\rightarrow\pi\pi$ amplitudes 
are obtained from the effective low-energy hamiltonian for $|\Delta S|=1$ 
transitions~[2\,-\,4],
\begin{equation}
{\cal H}_{ef\hspace{-0.5mm}f}^{\SS \Delta S=1}=\frac{G_F}{\sqrt{2}}
\;\xi_u\sum_{i=1}^8 c_i(\mu)Q_i(\mu)\hspace{1cm}(\mu < m_c)\,,
\label{ham}
\end{equation}
\begin{equation}
c_i(\mu)=z_i(\mu)+\tau y_i(\mu)\;,\hspace*{1cm}\tau=-\xi_t/\xi_u\;,
\hspace*{1cm}\xi_q=V_{qs}^*V_{qd}^{}\;.
\end{equation}
The arbitrary renormalization scale $\mu$ separates short- and
long-distance contributions to the decay amplitudes. The Wilson 
coefficient functions $c_i(\mu)$ contain all the information on heavy-mass 
scales. For CP conserving processes only the $z_i$ are numerically relevant.
The coefficient functions can be calculated for a scale $\mu \gtrsim 1\,$GeV 
using perturbative renormalization group techniques. They were computed in 
an extensive next-to-leading logarithm analysis by two groups
\cite{BJM,CFMR}. The local four-quark operators $Q_i(\mu)$ can be written, 
after Fierz reordering, in terms of color singlet quark bilinears:
\begin{eqnarray}
Q_1 &=& 4\,\bar{s}_L\gamma^\mu d_L\,\,\bar{u}_L\gamma_\mu u_L\,, 
\hspace*{2.49cm} 
Q_2 \,\,\,=\,\,\,\,4\,\bar{s}_L\gamma^\mu u_L\,\,\bar{u}_L
\gamma_\mu d_L\,, \label{qia} \\[4mm] 
Q_3 &=& \,4\,\sum_q \bar{s}_L\gamma^\mu d_L\,\,\bar{q}_L\gamma_\mu q_L\,,  
\hspace*{1.76cm}
Q_4 \,\,\,=\,\,\, \,4\,\sum_q \bar{s}_L\gamma^\mu q_L\,\,\bar{q}_L
\gamma_\mu d_L\,,\\[2mm]
Q_5 &=& \,4\,\sum_q \bar{s}_L\gamma^\mu d_L\,\,\bar{q}_R\gamma_\mu q_R\,, 
\hspace*{1.74cm}
Q_6 \,\,\,=\,\,\, \,-8\,\sum_q \bar{s}_L q_R\,\,\bar{q}_R d_L\,,\\[1mm]
Q_7 &=& \,4\,\sum_q \frac{3}{2}e_q\,\bar{s}_L\gamma^\mu d_L\,\,
\bar{q}_R \gamma_\mu q_R\,,\hspace*{1.0cm}
Q_8 \,\,\,=\,\,\, \,-8\,\sum_q \frac{3}{2}e_q\,\bar{s}_L q_R\,\,
\bar{q}_R d_L\,, \label{qio}
\end{eqnarray}
where the sum goes over the light flavors ($q=u,d,s$) and
\begin{equation}
q_{R,L}=\frac{1}{2}(1\pm\gamma_5)q\,,\hspace{1cm} 
e_q\,= (2/3,\,-1/3,\,-1/3)\,.
\end{equation}
$Q_3,\ldots,Q_6$ arise from QCD penguin diagrams involving a virtual $W$ and
a $c$ or $t$ quark, with gluons connecting the virtual heavy quark to light
quarks. They transform as $(8_L,1_R)$ under $SU(3)_L\times SU(3)_R$ and 
solely contribute to $\Delta I=1/2$ transitions. $Q_7$ and $Q_8$ are 
electroweak penguin operators \cite{BW,BG1} which are less important for 
the $\Delta I=1/2$ rule. Long-distance contributions to the amplitudes $A_I$ 
are contained in the hadronic matrix elements of the four-quark operators,
\begin{equation}
\langle Q_i(\mu)\rangle_I \,\equiv\,\langle\pi\pi,\,I|\,Q_i(\mu)\,
|K^0\rangle\,,
\end{equation}
which are related to the $\pi^+\pi^-$ and $\pi^0\pi^0$ final states through
the isospin decomposition
\begin{eqnarray}
\langle Q_i\rangle_0&=&\frac{1}{\sqrt{6}}\left( 2\langle\pi^+\pi^-
|\,Q_i\,|K^0\rangle+\langle\pi^0\pi^0|\,Q_i\,|K^0\rangle\right)\,,\\[1mm]
\langle Q_i\rangle_2&=&\frac{1}{\sqrt{3}}\left( \langle\pi^+\pi^-
|\,Q_i\,|K^0\rangle-\langle\pi^0\pi^0|\,Q_i\,|K^0\rangle\right)
\,\,\,=\,\,\,\sqrt{\frac{2}{3}}\langle\pi^+\pi^0|\,Q_i\,|K^+\rangle\,.
\end{eqnarray}
They are difficult to calculate but can be estimated using non-perturbative 
techniques generally for $\mu$ around a scale of $1\,$GeV. 

Major progress in the understanding of the $\Delta I=1/2$ rule was made 
when it was observed that the short-distance (quark) evolution, which is 
represented by the Wilson coefficient functions in the effective 
hamiltonian of Eq.~(\ref{ham}), leads to both an enhancement of the $I=0$
and a suppression of the $I=2$ final state. The {\it octet enhancement}
\cite{GLAM} in the $(Q_1,Q_2)$ sector is dominated by the increase of $z_2$ 
when $\mu$ evolves from $M_W$ down to $\mu \simeq 1\,$GeV, whereas the 
suppression of the $\Delta I=3/2$ transition results from a partial 
cancellation between the contributions from the $Q_1$ and $Q_2$ operators 
owing to a destructive Pauli interference in the $K^+\rightarrow\pi^+\pi^0$ 
amplitude. Another important short-distance enhancement was found to arise 
in the sector of the QCD penguin operators, in particular for $z_6$, through 
the proper inclusion of the threshold effects (and the associated incomplete 
GIM cancellation above the charm quark mass) \cite{BBG2}. Nevertheless,
it was concluded that the perturbative QCD effects are far from sufficient 
to describe the $\Delta I=1/2$ rule and QCD dynamics at low energies must 
be addressed. The long-distance enhancement of the matrix elements of the 
QCD penguin operators over the matrix elements of $Q_1$ and $Q_2$ was first 
conjectured and estimated in Ref.~\cite{VSZ} in the vacuum saturation 
approximation (VSA) \cite{VSA}. The VSA approach, however, fails completely 
in explaining the $\Delta I=1/2$ rule, and a more refined method for the 
calculation of the hadronic matrix elements~is~ certainly~needed. 

Due to the non-perturbative nature of the long-distance contribution, 
a large variety of techniques has been proposed to estimate it (for some 
recent publications see Refs.~[11\,-\,16]). Among the analytical methods, 
the $1/N_c$ expansion \cite{thoo} ($N_c$ being the number of colors)  
associated with the effective chiral lagrangian is particularly interesting. 
In this approach, QCD dynamics at low energies is represented by the `meson 
evolution' of the operators, from zero momentum to $\mu$, in terms of the 
chiral loop corrections to the matrix elements \cite{BBG2,BBG3}. The authors 
of Ref.~\cite{BBG3} calculated the loop corrections to the matrix elements of 
$Q_1$ and $Q_2$ and included the gluon penguin operator $Q_6$ at the tree 
level, consistent with the $1/N_c$ expansion. They obtained an additional 
enhancement and suppression of the $\Delta I=1/2$ and $\Delta I=3/2$ 
amplitudes, respectively, systematically continuing the octet enhancement 
in the $(Q_1,Q_2)$ sector to the long-distance domain. Numerically, $a_2$ 
was reproduced with an accuracy of 70 to approximately 100$\,\%$, whereas 
$a_0$ [for $\Lambda_{\mbox{\tiny QCD}}=300\,\mbox{MeV}$ and
$m_s(1\,\mbox{GeV})=125\,$-$\,175\,$MeV] was found to be around 
$65\,$-$\,80\,\%$ of the measured value, suggesting that the bulk of the
physics behind the $\Delta I=1/2$ rule in kaon decays is now understood. 
One might note that the agreement with experiment is not improved by 
including the next-to-leading order values for the $z_i$ \cite{BBL}.

In this article we present a new calculation of the hadronic matrix elements
in $K\rightarrow\pi\pi$ decays in the $1/N_c$ expansion for pseudoscalar
mesons. The paper contains several improvements over the original approach 
of Ref.~\cite{BBG3} which are conceptually and numerically important.
One improvement concerns the matching of short- and long-distance 
contributions to the amplitudes, by adopting a modified identification 
of virtual momenta in the integrals of the chiral loops. To be explicit, we 
consider the two currents or densities in the chiral representation of the
operators to be connected to each other through the exchange of an effective 
color singlet boson, and identify its momentum with the loop integration 
variable. The effect of this procedure is to modify the loop integrals, 
which introduces noticeable effects in the final results. More important 
it provides an unambiguous matching of the $1/N_c$ expansion in terms of 
mesons to the QCD expansion in terms of quarks and gluons. The approach 
followed here leads to an explicit classification of the diagrams into 
factorizable and non-factorizable. Factorizable loop diagrams refer to 
the strong sector of the theory and give corrections whose scale dependence 
is absorbed in the renormalization of the chiral effective lagrangian. The 
non-factorizable loop diagrams have to be matched to the Wilson coefficients 
and should cancel scale dependences which arise from the short-distance 
expansion. In a recent publication together with W.A. Bardeen and 
E.A. Paschos \cite{HKPSB} we used this method to calculate the hadronic 
matrix elements of $Q_6$ and $Q_8$ which dominate the ratio $\directcp$.
In this paper we focus on the CP conserving amplitudes which, to a large
extent, are governed by the current-current operators $Q_1$ and $Q_2$.

In Ref.~\cite{BBG3} a mass scale replacing the complete dependence of the 
exact expressions on the meson masses was introduced in the chiral logarithms. 
Another improvement of this paper is that we investigate the exact expressions 
for the matrix elements using the matching prescription discussed above, i.e.,
we evaluate the complete finite terms from the non-factorizable diagrams. 
Moreover, we calculate the whole of the matrix elements, that is to say, we 
also take into account the subleading penguin operators. For consistency 
with Ref.~\cite{HKPSB} we also include the small effects of the singlet 
$\eta_0$. In the numerical analysis we take special care to separate the 
different contributions. In particular, we discuss the effect of the final 
state interaction phases which were not taken into account in 
Ref.~\cite{BBG3}. Uncertainties arising from the short-distance part of
the calculation are estimated by comparing the amplitudes obtained from 
the LO and the NLO Wilson coefficients, respectively. Finally, we also 
investigate the size of higher order corrections to the hadronic matrix
elements to critically examine the stability of our results within the 
pseudoscalar approximation.

In the second part of this work we investigate the matrix element
of the ($|\Delta S|=2$) $K^0-\bar{K}^0$ amplitude in the $1/N_c$ 
expansion following the same lines of thought. The introduction to
this calculation we postpone to the beginning of Section~5. Our results 
for the $K\rightarrow\pi\pi$ matrix elements were already discussed in 
part in Refs.~\cite{thom,ger}. For a more detailed presentation of the 
general method we refer the reader to Refs.~\cite{HKPSB,TP}.
 
The paper is organized as follows. In Section~2 we review the general
framework of the effective low-energy calculation and discuss the matching
of short- and long-distance contributions to the decay amplitudes. Then, 
in Section~3 we investigate the $K\rightarrow\pi\pi$~matrix elements. 
We show explicitly that the scale dependence of the factorizable loop 
diagrams is absorbed in the renormalization of the bare couplings, the 
meson wave functions and masses. We next calculate the non-factorizable 
loop corrections in the cutoff regularization scheme. In Section~4 we 
match them to the Wilson coefficients to obtain the isospin amplitudes. 
In Section~5 we extend the analysis to the ($|\Delta S|=2$) $K^0-\bar{K}^0$ 
transition. We compute the matrix element and match it to the short-distance 
coefficient function to determine the $\hat{B}_K$ parameter. In both sections 
we present our numerical results and compare them with those of the existing 
analyses. The conclusions can be found in Section~6.
%
%
\section{General Framework \label{gen}}
Following the lines of Ref.~\cite{HKPSB} we calculate the hadronic matrix 
elements of the local four-quark operators (with $|\Delta S|=1$, 2) in the 
$1/N_c$ expansion. To this end we start from the chiral effective 
lagrangian for pseudoscalar mesons which involves an expansion in momenta 
where terms up to ${\cal O}(p^4)$ are included \cite{GaL}. Keeping only 
terms of ${\cal O}(p^4)$ which contribute to the $K\rightarrow \pi\pi$ 
or the $K^0-\bar{K}^0$ matrix elements and are leading in $N_c$ it 
reads:\footnote{One might note that the mass term $\propto L_8$ contributes 
only to the matrix elements of $Q_6$ and $Q_8$ which were computed in 
Ref.~\cite{HKPSB}. Here we include it for completeness.}
\begin{eqnarray}
{\cal L}_{ef\hspace{-0.5mm}f}&=&\frac{f^2}{4}\Big(
\langle D_\mu U^\dagger D^{\mu}U\rangle
+\frac{\alpha}{4N_c}\langle \ln U^\dagger -\ln U\rangle^2
+r\langle {\cal M} U^\dagger+U{\cal M}^\dagger\rangle\Big) \nonumber\\[1mm] 
&& +rL_5\langle D_\mu U^\dagger D^\mu U({\cal M}^\dagger U
+U^\dagger{\cal M})\rangle+r^2L_8\langle {\cal M}^\dagger U{\cal M}^\dagger U
+{\cal M} U^\dagger{\cal M} U^\dagger \rangle \label{lagr}\,,
\end{eqnarray}
with $D_\mu U=\partial_\mu U-ir_\mu U+iUl_\mu$, $\langle A\rangle$ denoting 
the trace of $A$ and ${\cal M}=\mbox{diag}( m_u,m_d,m_s)$. $l_\mu$~and~$r_\mu$
are left- and right-handed gauge fields, respectively, $f$ and $r$ are 
free parameters related to the pion decay constant $F_\pi$ and to the quark 
condensate, with $r = - 2 \langle \bar{q}q\rangle/f^2$. The complex matrix 
$U$ is a non-linear representation of the pseudoscalar meson nonet:
\begin{equation}
U=\exp\frac{i}{f}\Pi\,,\hspace{1cm} \Pi=\pi^a\lambda_a\,,\hspace{1cm} 
\langle\lambda_a\lambda_b\rangle=2\delta_{ab}\,, 
\end{equation}
where, in terms of the physical states
\begin{equation}
\Pi=\left(
\begin{array}{ccc}
\T\pi^0+\frac{1}{\sqrt{3}}a\eta+\sqrt{\frac{2}{3}}b\eta'
& \sqrt2\pi^+ & \sqrt2 K^+  \\[2mm]
\sqrt2 \pi^- & \T
-\pi^0+\frac{1}{\sqrt{3}}a\eta+\sqrt{\frac{2}{3}}b\eta' & \sqrt2 K^0 \\[2mm]
\sqrt2 K^- & \sqrt2 \bar{K}^0 & 
\T -\frac{2}{\sqrt{3}}b\eta+\sqrt{\frac{2}{3}}a\eta'
\end{array} \right)\,,
\end{equation}
with
\begin{equation}
a= \cos \theta-\sqrt{2}\sin\theta\,, \hspace{1cm}
\sqrt{2}b=\sin\theta+\sqrt{2}\cos\theta\,, \label{abth}
\end{equation}
The various conventions and definitions we use are in agreement with
Ref.~\cite{HKPSB}. In particular, we introduce the singlet $\eta_0$
in the same way and with the same value for the $U_A(1)$ symmetry breaking
parameter, $\alpha=\me^2+\mep^2-2\mk^2\simeq 0.72\,\mbox{GeV}^2$,
corresponding to the $\eta-\eta'$ mixing angle $\theta = -19^\circ$
\cite{eta}. The bosonic representation of the quark currents is defined 
in terms of (functional) derivatives of the chiral action:
\begin{eqnarray}
\bar{q}_{iL}\gamma^\mu q_{jL}&\hspace*{-1mm}\equiv& 
\hspace*{-1mm}\frac{\delta S}{\delta(l_\mu(x))_{ij}}\,=\,
-i\frac{f^2}{2}\big(U^\dagger\partial^\mu U\big)_{ji} \nonumber\\[2mm]
&&\hspace*{-1mm} 
+irL_5\big(\partial^\mu U^\dagger{\cal M}-{\cal M}^\dagger\partial^\mu U
+\partial^\mu U^\dagger U{\cal M}^\dagger U
-U^\dagger{\cal M} U^\dagger\partial^\mu U\big)_{ji}\,,\label{curr}
\end{eqnarray}
and the right-handed currents are obtained by parity transformation.
Eq.~(\ref{curr}) allows us to express the current-current operators 
in terms of the pseudoscalar meson fields.

The $1/N_c$ corrections to the matrix elements $\langle Q_i\rangle_I$ are
calculated by chiral loop diagrams in line with Ref.~\cite{HKPSB}. The
factorizable contributions, on the one hand, refer to the strong sector of
the theory and give corrections whose scale dependence is absorbed in the 
renormalization of the chiral effective lagrangian. This property is
obvious in the case of the (conserved) currents and was demonstrated 
explicitly in the case of the bosonized densities \cite{HKPSB,TP}. 
Consequently, the factorizable loop corrections can be computed within 
dimensional regularization. The non-factorizable corrections, on the other 
hand, are UV divergent and must be matched to the short-distance part. They 
are regularized by a finite cutoff which is identified with the short-distance
renormalization scale \cite{BBG3,BBL,BBG4,DO1}. The definition of the 
momenta in the loop diagrams which are not momentum translation invariant 
was discussed in detail in Ref.~\cite{HKPSB}. A consistent matching is 
obtained by considering the two currents or densities to be connected to 
each other through the exchange of a color singlet boson and by assigning 
the same momentum to it at long and short distances~[28\,-\,31]. 
The identification of this momentum with the loop integration variable 
leads to modified integrals in the chiral loop diagrams compared to those
of Refs.~\cite{BBG3,BBG4}. The numerical implications for the isospin 
amplitudes in $K\rightarrow\pi\pi$ decays and the $\hat{B}_K$ parameter 
will be addressed in Sections~4 and~5.

In this paper we investigate the hadronic matrix elements at leading and 
next-to-leading order in the chiral and the $1/N_c$ expansions. In
particular, we calculate the ${\cal O}(p^2/N_c)$ corrections to the 
current-current operators, that is to say, the one-loop corrections 
over the ${\cal O}(p^2)$ lagrangian. The matrix elements of the
density-density operators $Q_6$ and $Q_8$ are taken from Ref.~\cite{HKPSB}.
In the numerical analysis of the $\Delta I=1/2$ rule and the $\hat{B}_K$ 
parameter we use the leading logarithmic (LO), as well as, the 
next-to-leading logarithmic (NLO) values \cite{BJM,CFMR,BJW,HN} for the 
($|\Delta S|=1,$ 2) short-distance coefficient functions.\footnote{We treat
the coefficient functions as leading order in $1/N_c$ since the large 
logarithms arising from the long renormalization group evolution 
from $(m_t,M_W)$ to $\mu\simeq {\cal O}(1\,\mbox{GeV})$ compensate for the 
$1/N_c$ suppression.} In general, the lack of any reference to the 
renormalization scheme dependence in the effective low-energy calculation 
prevents a complete matching at the next-to-leading order~\cite{AJB98}. 
Nevertheless, a comparison of the amplitudes obtained from the LO and  
NLO coefficients is meaningful in order to test the validity of perturbation
theory.

In the following sections we calculate the long-distance $1/N_c$ corrections 
to the $K\rightarrow\pi\pi$ amplitudes and the $\hat{B}_{K}$ parameter. First,
we investigate the factorizable corrections~and show their absorption in the 
low-energy constants. Secondly, we determine the non-factorizable loops 
within the modified momentum prescription. Finally, we perform a numerical 
analysis and compare our results with those of the existing studies.
%
%
\section{\boldmath $K\rightarrow\pi\pi$ \unboldmath Decays}
In this section we present the hadronic matrix elements of the
current-current operators for the physical decay modes $K^0\rightarrow
\pi^+\pi^-$ and $K^0\rightarrow\pi^0\pi^0$ up to ${\cal O}(p^4)$ and 
${\cal O}(p^2/N_c)$ in the parameter expansion. From these results we 
derive the isospin amplitudes $K\rightarrow (\pi\pi)_{I=0,2}$, heading 
for an explanation of the $\Delta I = 1/2$ selection rule in kaon decays. 
%
%
\subsection{Factorizable \boldmath $1/N_c$ \unboldmath Corrections}
\begin{figure}[t]
\centerline{\epsfig{file=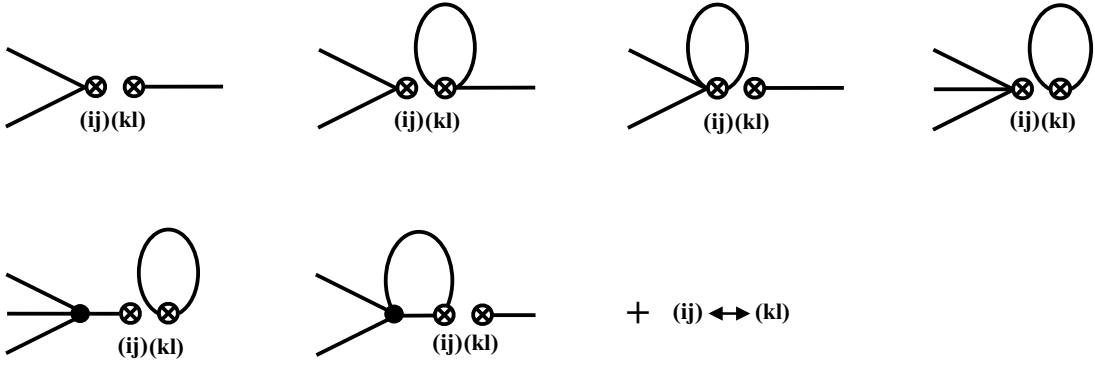,width=14.5cm}}
\vspace*{3.9mm}
\noindent
\caption{
Factorizable diagrams for the matrix elements of the current-current
operators in the isospin limit. Crossed circles represent the bosonized 
currents, black circles the strong vertices. The lines denote the 
pseudoscalar mesons. The external legs represent all possible permutations 
of the kaon and the pions.\label{ccf}}
\end{figure}

The (bare) tree level of the $K\rightarrow\pi\pi$ matrix elements, up to
${\cal O}(p^4)$ in the chiral expansion, as well as, the factorizable $1/N_c$
corrections to the ${\cal O}(p^2)$ can be calculated from the tree and loop
topologies depicted in Fig.~\ref{ccf}. From the sum of these diagrams we
obtain\footnote{In distinction to Ref.~\cite{HKPSB} the factor $i$ referring
to the weak vertex is included in the definition of the matrix element.} 
\begin{eqnarray}
\langle \pi^{+} \pi^{-}| Q_{2}| K^{0} \rangle^{F}_{(0)}
&=&\sqrt{2}\,f\,\left( \mk^2 - \mp^2 \right)\,
      \left[ 1 + \frac{4\,L_{5}}{f^2}\, \left( \mk^2 + 4\,\mp^2 \right) \right. \nonumber \\
&&\hspace{1cm}- \left. \frac{1}{16 \pi^2 f^2}\,
	\left( 3\,\lambda_{c}^2 -\frac{5}{4} \left( \mk^2 + 2 \mp^2 \right) 
        \,\log \lambda_{c}^2 \right) + \cdots \right]
\label{fdq2knppm} \,,
\end{eqnarray}
where
\begin{eqnarray}
\langle \pi^{+} \pi^{-}| Q_{2}| K^{0} \rangle^{F}
&=& \langle \pi^{+} \pi^{-}| Q_{4}| K^{0} \rangle^{F}
\,\,\,=\,\,\,-\,\langle \pi^{0} \pi^{0}| Q_{1}| K^{0} \rangle^{F} \nonumber
\\[1mm]
&=& \langle \hspace{0.5mm} \pi^{0} \hspace{0.5mm} \pi^{0} \hspace{0.5mm}| Q_{4}| K^{0} \rangle^{F}
\,\,\,=\,\,\,\frac{2}{3}\,\langle \pi^{0} \pi^{0}| Q_{7}| K^{0} \rangle^{F}
\label{fd-rest1}\,,
\end{eqnarray}
and
\begin{eqnarray}
\langle \pi^{+} \pi^{-}| Q_{i}| K^{0} \rangle^{F} 
&=&0\hspace{1cm}\mbox{for}\hspace{0.5cm}i\,\in\,\{1,3,5,7\} \\[2mm]
\langle \hspace{0.5mm}\pi^{0} \hspace{0.5mm}\pi^{0} 
\hspace{0.5mm}| Q_{i}| K^{0} \rangle^{F} 
&=&0\hspace{1cm}\mbox{for}\hspace{0.5cm}i\,\in\,\{2,3,5\} \,.
\label{fd-rest3}
\end{eqnarray}
The ellipses in Eq.~(\ref{fdq2knppm}) denote finite terms we omit here for
the analysis of the ultraviolet behaviour (in particular, they provide the 
reference scale for the logarithms). We specify our results in the cutoff 
regularization scheme to demonstrate the absorption of the quadratic, as 
well as, the logarithmic divergences as required by current conservation.
We note that all factorizable terms quadratic and logarithmic in the cutoff
are independent of the momentum prescription in the loop. $\lambda_c$ is the 
cutoff for the factorizable diagrams. We introduce two different scales since 
the factorizable and the non-factorizable corrections refer to disconnected 
sectors of the theory (strong and weak sectors). Having demonstrated the
absence of UV divergent terms in the sum of the factorizable diagrams, 
in the numerical analysis of the full expressions we will use dimensional 
regularization, as in pure chiral perturbation theory, which is momentum 
translation invariant. 
\noindent

If we renormalize the wave functions of the kaon and the pions $(\pi_r
\equiv Z_\pi^{1/2}\pi_0)$, as well as, the bare decay constant $f$ by 
using Eqs.~(14)-(17) and (25) of Ref.~\cite{HKPSB}, we arrive at the 
renormalized (factorizable) matrix elements of the ($|\Delta S|=1$) 
current-current operators:\footnote{The full expressions for the wave 
function and the decay constants are given in terms of integrals in 
Appendix A of Ref.~\cite{HKPSB}.}
\begin{equation}
\langle \pi^{+} \pi^{-}| Q_{2}| K^{0} \rangle^{F}_{(r)}
\,=\,\sqrt{2}\,F_{\pi}\,\left( \mk^2 - \mp^2 \right) \,
\left[ 1 + \frac{4\,\hat{L}_{5}^{r}}{F_{\pi}^2}\,\mp^2 \right] 
\label{fdq2knppmr} \,,
\end{equation}
where the constant $\hat{L_5^r}$ is defined through the relation
\cite{HKPSB}
\begin{equation} 
\frac{F_K}{F_\pi}\equiv 1+\frac{4\hat{L}_5^r}{F_\pi^2}(\mk^2-\mp^2)\,,
\label{kp1}
\end{equation}
and the remaining matrix elements can be obtained from 
Eqs.~(\ref{fd-rest1})-(\ref{fd-rest3}).

We notice that for the four-quark operators $Q_i$ of the current-current
type the divergent terms are absorbed by the renormalization procedure. In
addition, the factorizable $1/N_c$ corrections vanish completely, that is 
to say, the divergent as well as the finite terms. This property has been 
observed numerically, within dimensional regularization, because the complexity 
of all factorizable contributions prevents us from doing a fully analytic 
calculation. Since the factorizable scale $\lambda_c$ disappears through
renormalization, the only matching between long- and short-distance
contributions is obtained by identifying the cutoff scale $\Lambda_c$ 
of the non-factorizable diagrams with the QCD renormalization scale.

Finally, we note that in the next-to-leading order term of 
Eqs.~(\ref{fdq2knppmr}) and (\ref{kp1}) we used $1/F_{\pi}$ rather than 
$1/f$ as it was done in Ref.~\cite{BBG3}. Formally, the difference 
represents higher order effects. Nevertheless, the appearance of $1/f$ 
gives rise to a residual dependence on the factorizable scale $\lambda_c$, 
which has no counterpart at the short-distance level and will be absorbed 
by factorizable loop corrections to the matrix elements at the next order 
in the parameter expansion. Consequently, it is a more adequate choice 
to use the physical decay constant in the expressions under consideration.
Instead of $F_\pi$ the kaon decay constant $F_K$ could be used as well. 
Both choices will be considered in the numerical analysis, which gives 
a rough estimate of higher order corrections. 
%
%
\subsection{Non-factorizable \boldmath $1/N_c$ \unboldmath Corrections}
\noindent
\begin{figure}[t]
\centerline{\epsfig{file=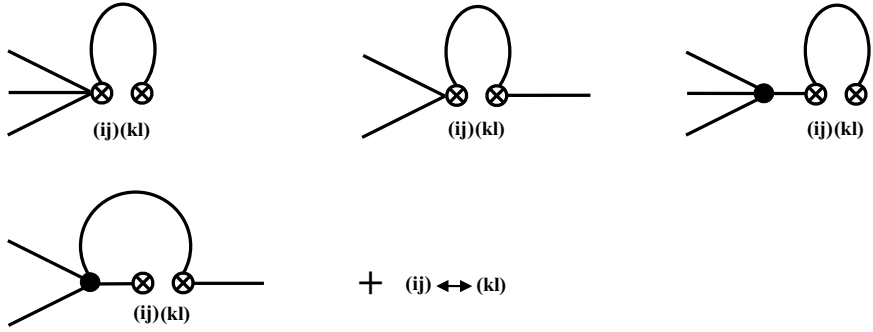,width=11.5cm}}
\vspace*{3.9mm}
\noindent
\caption{
Non-factorizable diagrams for the matrix elements of the current-current
operators in the isospin limit.\label{ccnf}}
\end{figure}
The non-factorizable $1/N_c$ corrections to the hadronic matrix elements
constitute the part to be matched to the short-distance Wilson coefficient 
functions; i.e., the corresponding scale $\Lambda_c$ has to be identified 
with the renormalization scale $\mu$ of QCD. We perform this identification, 
as we argued in Section 2, by associating the cutoff to the effective 
color singlet boson. Then, at the ${\cal O}(p^2)$ in the chiral expansion, 
from the diagrams of Fig.~\ref{ccnf} we obtain in the $SU(2)$ limit:
\begin{eqnarray}
\langle \pi^{+} \pi^{-}| Q_{1}| K^{0} \rangle^{\NF}
&=&-\frac{\sqrt{2}\,\left( \mk^2 - \mp^2 \right)}{16 \pi^2 F_{\pi}}\,
	\left[ 3\,\Lambda_{c}^{2}
- \left( \frac{1}{4}\,\mk^2 + 3\,\mp^2 \right)\,\Log + \cdots \right]\quad
\label{q1knppmnfd}\\[3mm]
\langle \pi^{+} \pi^{-}| Q_{2}| K^{0} \rangle^{\NF}
&=&\frac{\sqrt{2}\,\left( \mk^2 - \mp^2 \right)}{16 \pi^2 F_{\pi}}\,
	\left[ \frac{3}{2}\,\Lambda_{c}^{2}
+\left( \mk^2 - \frac{3}{2}\,\mp^2 \right)\,\Log + \cdots \right]\quad
\label{q2knppmnfd} \\[3mm]
\langle \pi^{+} \pi^{-}| Q_{3}| K^{0} \rangle^{\NF}
&=&\frac{\sqrt{2}\,\left( \mk^2 - \mp^2 \right)}{16 \pi^2 F_{\pi}}\,2
	\mp^2\,\Log + \cdots 
\label{q3knppmnfd} \\[3mm]
\langle \pi^{+} \pi^{-}| Q_{4}| K^{0} \rangle^{\NF}
&=&\frac{\sqrt{2}\,\left( \mk^2 - \mp^2 \right)}{16 \pi^2 F_{\pi}}\,
	\left[ \frac{9}{2}\,\Lambda_{c}^{2}
+ \left( \frac{3}{4}\,\mk^2 - \frac{5}{2}\,\mp^2 \right)\,\Log + \cdots \right]
\label{q4knppmnfd} \\[3mm]
\langle \pi^{+} \pi^{-}| Q_{7}| K^{0} \rangle^{\NF}
&=&\frac{\sqrt{2}\,\left( \mk^2 + 2\,\mp^2 \right)}{16 \pi^2 F_{\pi}}
\nonumber \\
&& \times \left[ \frac{9}{4}\,\Lambda_{c}^{2} - \frac{1}{8}\,\left( 
3\,\mk^2 + 7\,\mp^2 + \frac{6\,\mp^{4}}{\mk^{2}+2\,\mp^{2}}\right)\,\Log 
+ \cdots \right]
\label{q7knppmnfd} \\[3mm]
\langle \hspace{0.5mm} \pi^{0} \hspace{0.5mm} \pi^{0} \hspace{0.5mm}| Q_{2}| 
K^{0} \rangle^{\NF}
&=&\frac{\sqrt{2}\,\left( \mk^2 - \mp^2 \right)}{16 \pi^2 F_{\pi}}\,
	\left[ \frac{9}{2}\,\Lambda_{c}^{2}
+\frac{3}{4}\,\left( \mk^2 - 6\,\mp^2 \right)\,\Log + \cdots \right]
\label{q2knpnnnfd} \\[3mm]
\langle \hspace{0.5mm}\pi^{0} \hspace{0.5mm}\pi^{0} \hspace{0.5mm}| 
Q_{4}| K^{0} \rangle^{\NF}
&=&\frac{\sqrt{2}\,\left( \mk^2 - \mp^2 \right)}{16 \pi^2 F_{\pi}}\,
	\left[ \frac{9}{2}\,\Lambda_{c}^{2}
+\frac{1}{4}\,\left( 3\,\mk^2 - 10\,\mp^2 \right)\,\Log + \cdots \right]
\,,\hspace*{7mm} 
\label{q4knpnnnfd} 
\end{eqnarray}
where
\begin{eqnarray}
\langle \pi^{+} \pi^{-}| Q_{3}| K^{0} \rangle^{\NF}
&=&\langle \hspace{0.5mm}\pi^{0}\hspace{0.5mm} \pi^{0}\hspace{0.5mm}| Q_{3}| K^{0} \rangle^{\NF}
\,\,\,=\,\,\,- \langle \hspace{0.5mm}\pi^{0} \hspace{0.5mm}\pi^{0}\hspace{0.5mm}| Q_{5}| K^{0} \rangle^{\NF} 
\nonumber \\[2mm]
&=&\frac{1}{2}\langle \hspace{0.5mm}\pi^{0} \hspace{0.5mm}\pi^{0}\hspace{0.5mm}| Q_{7}| K^{0} \rangle^{\NF}
\,\,\,=\,\,\,- \langle \pi^{+} \pi^{-}| Q_{5}| K^{0} \rangle^{\NF} 
\label{nfd-rest2}
\end{eqnarray}
and
\begin{equation}
\langle \pi^{0} \pi^{0}| Q_{1}| K^{0} \rangle^{\NF}\,\,=\,\,0\,.
\label{nfd-rest4}
\end{equation}
One might note that in Eqs.~(\ref{q1knppmnfd})-(\ref{q4knpnnnfd}) [as in 
Eq.~(\ref{fdq2knppm})] we replaced $\me^2$, $\mep^2$, and the mixing angle 
$\theta$ by $\mp^2$ and $\mk^2$ using the octet-singlet mass matrix of
Ref.~\cite{eta}.

At this stage of the calculation we find quadratic, as well as, logarithmic 
divergences of the non-factorizable corrections. We note that already the
leading ($\sim \Lambda_c^2$) terms depend on the momentum prescription. The
quadratic terms were calculated in Ref.~\cite{FG} in the background field
method. In this paper we investigate the full expressions for the matrix
elements needed for the numerical analysis of the amplitudes. The results 
contain finite terms, originating from the solutions of the integrals listed 
in Eq.~(\ref{I4}) and in Appendix~B of Ref.~\cite{HKPSB}, which we neglect 
here for brevity and denote by the ellipses. We also note that in the case 
of $Q_7$ the solution of the integrals brings along a quartic dependence 
on the cutoff which has to be cancelled by adding a specific contact
interaction proportional to $\delta^{(4)}(0)$ to the Feynman rules of 
the truncated meson theory \cite{FG,GJLW}.

Even though the scale dependence of the perturbative coefficient functions
is only logarithmic, the full long-distance contribution including the 
quadratic terms has to be matched to the short-distance part. The quadratic 
dependence on the cutoff is physical and is necessary for several reasons.
First, in the chiral limit $(m_{q} = 0)$ all corrections vanish except for 
the $\lc^{2}$ terms, which bring in the only scale to be matched to the 
short distance. Secondly, they stabilize the $1/N_c$ expansion and generally 
improve the matching of the meson and the quark pictures \cite{BBG3}.
Finally, they provide us with a rough estimate of the contributions from 
higher resonances.

We note that in Eqs.~(\ref{q1knppmnfd})--(\ref{q4knpnnnfd}) we used the 
physical decay constant $F_{\pi}$ rather than $f$ in the same way as for 
the factorizable diagrams. Again the difference represents higher order 
effects. However, the (factorizable) scale dependence of $f$ has no 
counterpart in the short distance and will be absorbed at the next order 
in the chiral expansion. As for the factorizable contributions the choice 
of $F_K$ instead of $F_\pi$ would be also appropriate.
%
%
\section{Numerical Analysis}
In this paragraph we list the numerical values for the hadronic 
matrix elements. We next match them to the Wilson coefficients and 
study the $K\rightarrow(\pi\pi)_{I=0,\,2}$ isospin amplitudes. In 
Section~\ref{hme} we discuss in detail the $1/N_c$ corrections to the 
matrix elements. In this context we also calculate the bag parameters, 
which quantify the deviations from the results obtained in the vacuum 
saturation approximation and, therefore, are convenient for a comparison 
with other works. The main results of the present analysis can be found 
in Section~\ref{delta}. Therein we give the amplitudes $a_0$ and $a_2$ 
as functions of the matching scale and compare them with the data.
%
%
\subsection{Hadronic Matrix Elements \label{hme}}
Throughout the numerical analysis we use the following values for 
the parameters \cite{pdg98}:
\[
\begin{array}{lclcllcl}
m_\pi&\equiv&\,\big(m_{\pi^0}+m_{\pi^+}\big)/2\,&=& 
137.3\,\,\mbox{MeV}\,,\hspace{0.8cm}&F_\pi&=&92.4\,\,\mbox{MeV}\,,
\end{array}
\]
\[
\begin{array}{lclcllcl}
\mk  &\equiv& \big(m_{K^0}+m_{K^+}\big)/2 &=& 495.7\,\,\mbox{MeV}\,,
\hspace*{0.6cm}&F_K&=&113\,\,\,\mbox{MeV}\,,\\[1.0mm]
m_\eta   &=& 547.5\,\,\mbox{MeV}\,,&&&\theta&=&-19^\circ\,,
\\[1.0mm]
m_{\eta'}&=& 957.8\,\,\mbox{MeV}\,,&&&G_F&=&1.1664\cdot 10^{-5}\,\,
\mbox{GeV}^{-2}\,,\\[1.0mm]
|V_{ud}|&=&0.974\,,   &&& |V_{us}|&=&0.22\,.\\[1.0mm]
\end{array}
\]
Substituting them in Eq.~(\ref{kp1}) we compute $\hat{L}_5^r=2.07\times 
10^{-3}$. 

We parameterize our results in terms of the non-perturbative bag parameters 
$B_{i}^{(1/2)}$ and $B_{i}^{(3/2)}$, which quantify the deviations from 
the values obtained in the vacuum saturation approximation \cite{VSA}:
\begin{eqnarray}
B_{i}^{(1/2)}&=&\frac{\mbox{Re}\langle Q_{i} \rangle_{0}}
{\langle Q_{i} \rangle_{0}^{\scriptsize \mbox{VSA}}}\,,\quad i \, \in \, 
\{1,\ldots ,8\}\,,  \\[1mm]
B_{i}^{(3/2)}&=&\frac{\mbox{Re}\langle Q_{i} \rangle_{2}}
{\langle Q_{i} \rangle_{2}^{\scriptsize \mbox{VSA}}}\,,\quad i \, \in \, 
\{1,2,7,8\} \,,
\end{eqnarray}
with $\langle Q_{i}\rangle_{I}$ containing both factorizable and
non-factorizable contributions. The VSA expressions for the matrix elements 
are taken from Eqs.~(XIX.11)-(XIX.28) of Ref.~\cite{BBL}.\footnote{Note that 
our definition of the pion decay constant ($F_\pi=92.4\,$MeV) differs by a 
factor of $1/\sqrt{2}$ from the one used in Ref.~\cite{BBL}.} The numerical 
values for the matrix elements of the current-current operators are given 
in Tables~\ref{tab1} and~\ref{tab2}. $\langle Q_5\rangle_0^{\scriptsize
\mbox{VSA}}$ and $\langle Q_7\rangle_{0,2}^{\scriptsize \mbox{VSA}}$ are 
functions of $R\equiv 2\mk^2/(m_s+m_d)\simeq 2\mk^2/m_s$ and, consequently, 
depend on the renormalization scale. For comparison, in the tables we 
also show the results obtained in the large-$N_c$ limit, see 
Eqs.~(\ref{fd-rest1})-(\ref{fdq2knppmr}). One might note that the 
different values generally do not coincide, even if the small 
${\cal O}(p^4)$ term proportional to $m_\pi^2$ in Eq.~(\ref{fdq2knppmr}) 
[which contributes only at the level of $2\,\%$ of the ${\cal O}(p^2)$ 
tree level term] is neglected, since in the vacuum saturation 
approximation Fierz terms are taken into account which are 
subleading in $N_c$. In particular, the matrix element $\langle Q_1
\rangle_0^{\scriptsize\mbox{VSA}}$ differs by a factor of $(1-2/N_c)$ 
from the result obtained at the ${\cal O}(p^2)$ in the large-$N_c$ limit. 
We notice that the inclusion in part of the $1/N_c$ corrections in the VSA 
method leads to a suppression and enhancement of the $I=0$ and $I=2$ 
amplitudes, respectively, in complete disagreement with the data. 
\begin{table}[t]
\begin{eqnarray*}
\begin{array}{|c||c|c|c|c|c|l|}\hline
& \hspace*{1mm}\langle Q_1\rangle_0\hspace*{1mm} 
& \hspace*{1mm}\langle Q_2\rangle_0\hspace*{1mm} 
& \hspace*{1mm}\langle Q_3\rangle_0\hspace*{1mm}
& \hspace*{1mm}\langle Q_4\rangle_0\hspace*{1mm} 
& \hspace*{1mm}\langle Q_5\rangle_0\hspace*{1mm} 
& \hspace*{10mm}\langle Q_7\rangle_0
\\ 
\hline\hline
\rule{0cm}{5.5mm}
\mbox{\small VSA}  & -4.03 & 20.2 & 12.1 & 36.3 & -11.7\cdot R^2 & 
18.2+32.5\cdot
R^2 \\[0.5mm] 
\mbox{\small tree} & -12.3 & 24.6 & 0    & 37.0 & 0 & 18.5  \\[1mm]
\hline
\end{array}
\end{eqnarray*}
\caption{
$I=0$ matrix elements of the current-current operators: VSA vs.~tree level 
(large-$N_c$ limit), in units of $10^{6} \cdot \mbox{MeV}^{3}$ ($R$ in units 
of GeV). \label{tab1}}
\end{table}
\begin{table}[tbh]
\begin{eqnarray*}
\begin{array}{|c||c|c|l|}\hline
& \hspace*{1mm}\langle Q_1\rangle_2\hspace*{1mm} 
& \hspace*{1mm}\langle Q_2\rangle_2\hspace*{1mm} 
& \hspace*{10mm}\langle Q_7\rangle_2
\\ 
\hline\hline
\rule{0cm}{5.5mm}
\mbox{\small VSA}  & 22.8 & 22.8 & -25.7+18.9\cdot R^2 \\[0.5mm] 
\mbox{\small tree} & 17.4 & 17.4 & -26.1  \\[1mm]
\hline
\end{array}
\end{eqnarray*}
\caption{Same as in Table~\ref{tab1}, now the $I=2$ matrix elements.
\label{tab2}}
\end{table}
\begin{table}[t]
\begin{eqnarray*}
\begin{array}{|c||c|c|c|c|c|c|c|}\hline
\lc&0.5\,\,\mbox{GeV}&0.6\,\,\mbox{GeV}&0.7\,\,\mbox{GeV}&0.8\,\,
\mbox{GeV}&0.9\,\,\mbox{GeV}&1.0\,\,\mbox{GeV}& \\ 
\hline\hline
\rule{0cm}{5mm}
\langle Q_{1} \rangle_{0} & -27.4 & -33.2 & -40.2 & -48.2 & -57.3 & -67.4  &  -5.55 i \\[0.5mm]
\langle Q_{2} \rangle_{0} & 50.0  & 58.8 & 68.8 & 79.9 & 92.4 & 106  & 11.1 i \\[0.5mm]
\langle Q_{3} \rangle_{0} & 0.04 & 0.05 & 0.03 & -0.02 & -0.12 & -0.26  & 0 \\[0.5mm]
\langle Q_{4} \rangle_{0} & 77.5 & 92.1 & 109 & 128 & 150 & 173  & 16.6 i \\[0.5mm]
\langle Q_{5} \rangle_{0} & -0.04 & -0.05 & -0.03 & 0.02 & 0.12 & 0.26  & 0 \\[0.5mm]
\langle Q_{6} \rangle_{0} & -44.1 & -38.6 & -33.7 & -29.4 & -25.5 & -21.9  & 0 \\[0.5mm]
\langle Q_{7} \rangle_{0} & 34.4 & 40.1 & 46.6 & 54.1 & 62.6 & 72.2  & 8.32 i \\[0.5mm]
\langle Q_{8} \rangle_{0} & 118 & 119 & 119 & 119 & 118 & 117 & 36.7 i \\[0.5mm]
\hline
\end{array}
\end{eqnarray*}
\caption{Hadronic matrix elements of $Q_{1,\ldots,5,7}$ (in units of $10^{6} 
\cdot \mbox{MeV}^{3}$) and $Q_{6,8}$ (in units of $R^2 \cdot \mbox{MeV}$)
in the isospin limit for the $I = 0$ amplitudes, shown for various values 
of the cutoff $\lc$.\label{tab3}}
\end{table}
\vspace*{1mm}
\begin{table}[t]
\begin{eqnarray*}
\begin{array}{|c||c|c|c|c|c|c|c|}\hline
\lc&0.5\,\,\mbox{GeV}&0.6\,\,\mbox{GeV}&0.7\,\,\mbox{GeV}&0.8\,\,\mbox{GeV}
&0.9\,\,\mbox{GeV}&1.0\,\,\mbox{GeV}& \\ 
\hline\hline 
\rule{0cm}{5mm}
\langle Q_{1} \rangle_{2} & 6.54  & 2.51 & -2.26 & -7.77 & -14.0 & -21.1  & -3.45 i \\[0.5mm]
\langle Q_{2} \rangle_{2} & 6.54  & 2.51 & -2.26 & -7.77 & -14.0 & -21.1  & -3.45 i \\[0.5mm]
\langle Q_{7} \rangle_{2} & -14.5  & -10.7 & -6.27 & -1.15 & 4.67 & 11.2  &  5.18 i \\[0.5mm]
\langle Q_{8} \rangle_{2} & 39.9  & 35.3 & 31.2 & 27.2 & 23.2 &  18.8 & -11.5 i \\[0.5mm]
\hline
\end{array}
\end{eqnarray*}
\caption{
Same as in Table~\ref{tab3}, now for the $I = 2$ amplitudes.\label{tab4}}
\end{table}

In Tables~\ref{tab3} and~\ref{tab4} we list our results for the hadronic 
matrix elements at next-to-leading order in the chiral and the $1/N_c$ 
expansions. The matrix elements of the current-current operators are 
calculated from Eqs.~(\ref{fd-rest1})-(\ref{nfd-rest4}) including the finite 
terms denoted by the ellipses. The results for the operators $Q_6$ and $Q_8$ 
are taken from Ref.~\cite{HKPSB}. These results contain the leading plus
next-to-leading order terms in the chiral expansion of the density-density
operators, as well as, the leading $1/N_c$ corrections, that is to say, 
the ${\cal O}(p^0)$, ${\cal O}(p^2)$, and ${\cal O}(p^0/N_c)$. Note that the 
matrix elements generally contain a non-vanishing imaginary part (scale
independent at the one-loop level) which is due to on-shell ($\pi-\pi$) 
rescattering~effects. 

The isospin amplitudes are largely dominated by the operators $Q_1$ and
$Q_2$. Therefore it is instructive to analyze in detail the $1/N_c$
corrections to these two operators. To this end we next give the 
analytic expressions for the isospin matrix elements of $Q_1$ and $Q_2$:
\begin{eqnarray}
\langle Q_1\rangle_0
&=& -\frac{1}{\sqrt{3}}F_\pi\left( \mk^2 - \mp^2 \right)
\left[1+\frac{4\hat{L}_5^r}{F_\pi^2}m_\pi^2+
\frac{1}{(4\pi)^2F_\pi^2}\right.
\nonumber\\
&&\times\left.\left( 6\Lambda_c^2-\Big(\frac{1}{2}\mk^2
+6m_\pi^2\Big)\log\Big(1+\frac{\Lambda_c^2}{\tilde{m}^2}\Big)
\right)\right]\,+\,a_{10}[\tilde{m}]  \label{isq10}\\[2mm]
\langle Q_2\rangle_0
&=& \frac{2}{\sqrt{3}}F_\pi\left( \mk^2 - \mp^2 \right)
\left[1+\frac{4\hat{L}_5^r}{F_\pi^2}m_\pi^2+
\frac{1}{(4\pi)^2F_\pi^2}\right.
\nonumber \\
&&\left.\times\left(\frac{15}{4}\Lambda_c^2+
\Big(\frac{11}{8}\mk^2-\frac{15}{4}m_\pi^2\Big)
\log\Big(1+\frac{\Lambda_c^2}{\tilde{m}^2}\Big)
\right)\right]\,+\,a_{20}[\tilde{m}]\hspace*{6mm} \\[2mm]
\langle Q_1\rangle_2\,\,\,=\,\,\,\langle Q_2\rangle_2
&=& \sqrt{\frac{2}{3}}F_\pi\left( \mk^2 - \mp^2 \right)
\left[1+\frac{4\hat{L}_5^r}{F_\pi^2}m_\pi^2+
\frac{1}{(4\pi)^2F_\pi^2}\right.
\nonumber\\
&&\times\left.\left( -3\Lambda_c^2+\Big(\frac{1}{4}\mk^2
+3m_\pi^2\Big)\log\Big(1+\frac{\Lambda_c^2}{\tilde{m}^2}\Big)
\right)\right]\,+\,a_{21}[\tilde{m}]\,. \label{isq22}
\end{eqnarray}
Eqs.~(\ref{isq10})-(\ref{isq22}) allow us to compare our results 
with the analytic expressions of Ref.~\cite{BBG3}. First, we note that the 
modified matching which was discussed in Section~2 increases the terms 
quadratic in the cutoff by a factor of 3/2 relative to the results presented 
therein. This was already observed in Ref.~\cite{FG}. The modification of 
the quadratic terms provides an additional octet enhancement in the 
long-distance domain. The logarithmic terms, on the other hand, are modified 
only on account of the presence of the $\eta_0$. To be explicit, in 
the octet limit [i.e., in the absence of the $\eta_0$, with $a=b=1$ and 
$m_\eta^2=(4\mk^2-m_\pi^2)/3$\,] the coefficient of the logarithm in 
Eq.~(\ref{isq10}) is reduced to $(\mk^2/2+10m_\pi^2/3)$ whereas the other 
terms remain unchanged. The separation of the logarithmic and the finite 
terms in Eqs.~(\ref{isq10})-(\ref{isq22}) is arbitrary and is done, for 
comparison with Ref.~\cite{BBG3}, by introducing a mass scale replacing 
the dependence of the exact expressions on the meson masses in the chiral 
logarithms. The logarithmic and the finite terms $(a_{iI})$ defined in 
this way each depend on the choice of the mass scale $\tilde{m}$, whereas 
the sum of all contributions is independent of this parameter. We 
calculated the complete finite terms arising from the non-factorizable 
loop diagrams using the matching prescription advocated in 
Refs.~\cite{HKPSB,FG}.\footnote{For details on the computation of the loop 
integrals see Appendix~B of Ref.~\cite{HKPSB}.} These terms were not included 
in Ref.~\cite{BBG3}. Consequently, the numerical values of the matrix 
elements reported therein exhibit a dependence on the specific choice of the
mass scale in the logarithms which is absent in the present calculation.

In Table~\ref{tab5} we split up the numerical values for the $I=0$ and $I=2$ 
matrix elements of $Q_1$ and $Q_2$ with respect to the quadratic, the 
logarithmic, and the finite terms, respectively, at a cutoff scale of 
$\Lambda_c =800\,$MeV. 
\begin{table}[t]
\begin{eqnarray*}
\begin{array}{|c||c|c|c|c|}\hline
& \langle Q_1\rangle_0 & \langle Q_1\rangle_2& \langle Q_2\rangle_0
& \langle Q_2\rangle_2
\\ 
\hline\hline
\rule{0cm}{5.5mm}
\mbox{\small tree} & -12.3 & 17.4 & 24.6 & 17.4 \\[0.5mm]
\T \Lambda_c^2 & -34.5 & -24.4 & 43.1 & -24.4  \\[0.5mm]
\T \log\Lambda_c[\tilde{m}] & 4.43 & 3.13 & 10.0 & 3.13 \\[0.5mm]
\mbox{\small finite} & \,-5.83-5.55i\, & \,-3.90-3.45i\, & 
\,\,\,2.20+11.1i\,\,\, & \,-3.90-3.45i\, \\[1mm]
\hline\hline 
\rule{0cm}{5.5mm}
\mbox{\small total} & -48.2-5.55i & -7.77-3.45i & 
79.9+11.1i & -7.77-3.45i\\[1mm]
\hline
\end{array}
\end{eqnarray*}
\caption{Different contributions to the hadronic matrix elements of 
$Q_1$ and $Q_2$ (in units of $10^{6} \cdot \mbox{MeV}^{3}$) for
$\Lambda_c=800\,$MeV and $\tilde{m}=300\,$MeV.\label{tab5}}
\end{table}
From the table we see that the finite terms are of the same order of magnitude 
as the logarithmic ones and, therefore, must be considered at the same level 
in the numerical analysis. These terms are generally suppressed  by a factor
of $\delta\equiv m_{K,\pi}^2/(4\pi F_\pi)^2<20\,\%$ with respect to the
leading ${\cal O}(p^2)$ tree level. In addition, as can be seen from 
Eqs.~(\ref{isq10})-(\ref{isq22}) and Table~\ref{tab5}, no coefficient larger 
than one or two which could significantly enhance them has been found. This 
is different from the quadratic terms which are not suppressed as their 
relative size is determined by $\Delta\equiv\Lambda_c^2/(4\pi F_\pi)^2$ 
and, moreover, they appear with larger prefactors [even as large as six in 
Eq.~(\ref{isq10})].\footnote{
It is interesting to note that the non-suppression of the quadratic terms 
presumably could be important for $Q_6$ but less important for $Q_8$. On 
the one hand, the first non-vanishing tree level contribution to the 
operators $Q_6$ and $Q_8$ is of the ${\cal O}(p^2)$ and ${\cal O}(p^0)$, 
respectively. On the other hand, the first non-vanishing quadratic 
corrections to both operators are of the ${\cal O}(p^2/N_c)$ (terms of 
the ${\cal O}(p^0/N_c)$ were found to be only logarithmic \cite{HKPSB}). 
Consequently, in the case of $Q_8$ the quadratic terms are (chirally)
suppressed by a factor of $p^2\hspace*{-0.4mm}\cdot\hspace*{-0.1mm}\Delta$ 
with respect to the (leading) tree level contribution whereas in the case 
of $Q_6$ they bring in only a factor of $\Delta$. Quadratic terms, even 
though subleading in $N_c$, could therefore significantly affect the matrix 
element of $Q_6$ especially if large prefactors are observed as for $Q_1$ 
and $Q_2$ in Eqs.~(\ref{isq10})-(\ref{isq22}). This difference between the 
$Q_6$ and $Q_8$ operators could play an important role for $\varepsilon'/
\varepsilon$. This point will be investigated in Ref.~\cite{HKPS}.}
Consequently, in the case of the $I=0$ matrix elements of $Q_1$ and $Q_2$ 
both the logarithmic and the finite corrections are moderate, and the chiral 
limit gives a satisfactory representation of the full amplitude provided
that the matching scale is taken sufficiently large ($\Lambda_c\gtrsim
500\,$-$\,600\,\mbox{MeV}$). In the case of the $I=2$ matrix elements we also 
observe that the quadratic terms are enhanced with respect to the tree level, 
whereas the logarithmic and the finite terms are largely suppressed. However, 
in this case the quadratic corrections counteract the tree level, and the sum 
of both contributions is no longer large compared to the logarithmic and the 
finite terms. Therefore the neglect of either of the terms is no longer 
justified. In particular, we observe that for the $\Delta I=3/2$ channel the 
chiral limit gives a better approximation to the exact result than a 
calculation which includes only the logarithms without taking into account 
the finite terms. This remark also holds for the matrix element $\langle 
Q_1\rangle_0$. Finally, we note that variation of the mass scale in the 
logarithms [$m_\pi<\tilde{m}<\mk]$ in Ref.~\cite{BBG3} has a noticeable 
effect on the numerical value of the $I=2$ amplitude. 

When comparing the results of the present analysis with those of
Ref.~\cite{BBG3} one has to take into account another difference 
in the treatment of the next-to-leading order terms: in 
Eqs.~(\ref{isq10})-(\ref{isq22}) we used $1/F_\pi$ rather than the bare
parameter $1/f$ as it was done in Ref.~\cite{BBG3}. Formally, the difference 
concerns higher order effects, as we already discussed above. However, since 
the factorizable scale which appears in the bare coupling $f$ will be 
absorbed by factorizable loop corrections to the matrix elements at the
next order in the parameter expansion, it has not to be matched to any 
short-distance contribution. Consequently, it is a more adequate choice 
to use the physical decay constant in the expressions under consideration. 
The effect of this different treatment of the next-to-leading order terms
will be further discussed in Section~\ref{delta}. 
 
\begin{table}[bht]
\begin{eqnarray*}
\begin{array}{|c||c|c|c|c|c|c|}\hline
\lc&0.5\,\,\mbox{GeV}&0.6\,\,\mbox{GeV}&0.7\,\,\mbox{GeV}&
0.8\,\,\mbox{GeV}&0.9\,\,\mbox{GeV}&1.0\,\,\mbox{GeV} \\ 
\hline\hline 
\rule{0cm}{5mm} 
B_{1}^{(1/2)}& 6.75 & 8.24 & 9.98 & 12.0 & 14.2 & 16.6  \\[0.5mm]
B_{2}^{(1/2)}& 2.47 & 2.91 & 3.41 & 3.96 & 4.57 &  5.23 \\[0.5mm]
B_{3}^{(1/2)}& 0.003 & 0.004 & 0.002 & -0.002 & -0.010 & -0.021 \\[0.5mm]
B_{4}^{(1/2)}& 2.12 & 2.54 & 3.00 & 3.53 & 4.13 & 4.75 \\[0.5mm]
B_{5}^{(1/2)}& 0.0004 & 0.0009 & 0.0005 & -0.0003 & -0.0014 & 
-0.0020 \\[0.5mm]
B_{6}^{(1/2)}& 1.26 & 1.10 & 0.96 & 0.84 & 0.72 & 0.62 \\[0.5mm]
B_{7}^{(1/2)}& 0.15 & 0.16 & 0.18 & 0.21 & 0.23 & 0.26 \\[0.5mm]
B_{8}^{(1/2)}& 1.20 & 1.21 & 1.21 & 1.21 & 1.20 & 1.19 \\[0.5mm]
\hline
\end{array}
\end{eqnarray*}
\caption{Bag parameters for the $I = 0$ amplitudes, shown for various values 
of the cutoff. $B^{(1/2)}_{5,\,7,\,8}$ depend on $R\simeq 2m_K^2/m_s$ and are 
calculated for a running $m_s(\mu=\lc)$ at the leading logarithmic order 
($\Lambda_{\mbox{\tiny QCD}}=325\,\mbox{MeV}$) with $m_s(1\,\mbox{GeV})=
175\,\mbox{MeV}$.\label{tab6}}
\end{table}
\begin{table}[htb]
\begin{eqnarray*}
\begin{array}{|c||c|c|c|c|c|c|}\hline
\lc&0.5\,\,\mbox{GeV}&0.6\,\,\mbox{GeV}&0.7\,\,\mbox{GeV}&
0.8\,\,\mbox{GeV}&0.9\,\,\mbox{GeV}&1.0\,\,\mbox{GeV} \\ 
\hline\hline 
\rule{0cm}{5mm}
B_{1}^{(3/2)}& 0.29  & 0.11 & -0.10 & -0.34 & -0.61 & -0.92 \\[0.5mm]
B_{2}^{(3/2)}& 0.29  & 0.11 & -0.10 & -0.34 & -0.61 & -0.92 \\[0.5mm]
B_{7}^{(3/2)}& -0.15  & -0.10 & -0.06 & -0.01 & 0.04 & 0.09  \\[0.5mm]
B_{8}^{(3/2)}&  0.72 & 0.64 & 0.56 & 0.49 & 0.42 & 0.34  \\[0.5mm]
\hline
\end{array}
\end{eqnarray*}
\caption{Same as in Table~\ref{tab6}, now for the $I = 2$ amplitudes. 
\label{tab7}}
\end{table}
In Tables~\ref{tab6} and~\ref{tab7} we list the values we compute for 
the bag parameters $B_i^{(1/2)}$ and $B_i^{(3/2)}$. We find a large 
enhancement of $B_{1}^{(1/2)}$ and $B_{2}^{(1/2)}$ over the VSA result, 
which constitutes the dominant contribution, at long distances,
to the $\Delta I = 1/2$ transition in $K\rightarrow\pi\pi$ decays. 
Moreover, we obtain the correct scale dependence counteracting the scale 
behaviour of the Wilson coefficients $z_{1}$ and $z_{2}$, which leads to an 
acceptable matching (see Section~\ref{delta}). In view of the large 
corrections one might question the convergence of the $1/N_c$ expansion. 
However, there is no strong reason for such doubts because the 
non-factorizable contribution we consider in this paper represents 
the first term in a new type of a series absent in the large-$N_c$ limit. 
It is reasonable to assume that this leading non-factorizable term carries 
a large fraction of the whole contribution \cite{BBG3} (see also the
discussion in Section~\ref{delta}). $B_3^{(1/2)}$ and 
$B_5^{(1/2)}$ turn out to be very close to zero. This property is due to 
the vanishing tree level, as well as, to the small $1/N_c$ corrections 
proportional to $m_\pi^2/(4\pi F_\pi)^2$, see Eqs.~(\ref{q3knppmnfd}) and 
(\ref{nfd-rest2}). We notice that the small contribution of the operator 
$Q_5$ to $\varepsilon'/\varepsilon$ is even further reduced when replacing 
the VSA expression for $\langle Q_5\rangle_0$, which is commonly used in
the analysis of $\varepsilon'/\varepsilon$ \cite{AJB98}, by the result
presented in this paper. $B_7^{(1/2)}$ and $B_7^{(3/2)}$ are also found to 
be significantly reduced with respect to vacuum saturation approximation.
In particular, $B_7^{(3/2)}$ turns out to be negative for small values
of the cutoff.\footnote{Very recently \cite{KPdR} the first non-trivial 
$1/N_c$ corrections to the matrix elements of $Q_7$ were evaluated using 
the methods of Ref.~\cite{KPR-PPR}. The numerical results were also sensitive
to the choice of the renormalization scale. In particular, negative values 
for $B_7^{(1/2)}$ and $B_7^{(3/2)}$ were found below $\mu\lesssim 1.3\,$GeV, 
in qualitative agreement with the results of the present analysis but in
disagreement with the large positive values obtained in the chiral 
quark model at a matching scale of $0.8\,$GeV \cite{BEF}.}
We also notice a decrease of the $B_{1}^{(3/2)}$ and $B_{2}^{(3/2)}$ 
parameters, which are relevant for $A_{2}$. However, as we will see below, 
their scale dependence largely overcompensates for the variation of the 
short-distance coefficient functions. Nevertheless, as the values are 
found to be reduced, they generally account for the reduction of the $I=2$ 
amplitude. Finally, $B_{6}^{(1/2)}$ receives only small corrections whereas 
$B_{8}^{(3/2)}$ comes out to be substantially reduced relative to the VSA 
result \cite{HKPSB}. The numerical implications for $\varepsilon'/\varepsilon$
will be investigated elsewhere \cite{HKPS}. One might note that the numerical
values of $B_{8}^{(3/2)}$ shown in Table~\ref{tab7} differ from the ones given 
in Table~2 of Ref.~\cite{HKPSB}. This is due to the fact that in the present 
paper we include only the real part of the hadronic matrix elements in the 
definition of the $B_i$ parameters (see Section~\ref{delta}). 
%
%
\subsection{The $\Delta I = 1/2$ Rule \label{delta}}
We next investigate the CP conserving amplitudes $\mbox{Re}a_{0}$ and 
$\mbox{Re}a_{2}$. 
To this end we start from the expression for the isospin amplitudes $A_I$ 
which contain the ($\pi-\pi$) strong interaction phase shift for the $I=0$ 
and the $I=2$ final states, respectively,
\begin{equation}
A_{I=0,2} \, = \,\frac{G_F}{\sqrt{2}}\,V_{ud}^{}
V_{us}^{*}\sum_{i}\,c_{i}(\mu)\, \langle Q_{i}(\mu) \rangle_{I=0,2}\,.
\label{amp1}
\end{equation}
Then
\begin{equation}
\mbox{Re} a_I \, = \, \frac{G_F}{\sqrt{2}}\,V_{ud}^{}
V_{us}^{*}\,\,\Big|\sum_{i}\,z_{i}\,\langle Q_{i}\rangle_I\Big|\,
=\,\frac{G_F}{\sqrt{2}}\,V_{ud}^{}V_{us}^{*}\,\frac{1}{\cos\delta_I}
\sum_{i}\,z_{i}\,\mbox{Re}\langle Q_{i}\rangle_I\,.
\label{amp2}
\end{equation}
Within an exact realization of non-perturbative QCD the two expressions in 
Eq.~(\ref{amp2}) are equivalent. However, in the approximate low-energy 
calculation of the present work the long-distance imaginary part which we 
computed at the one-loop level (see Tables~\ref{tab3} and~\ref{tab4}) is not 
expected to be of the same accuracy as the real part obtained at this level. 
In particular, as the one-loop (long-distance) imaginary part is scale 
independent, it cannot compensate for the scale dependence of the Wilson 
coefficients $z_i$ leading to a scale dependent imaginary part of the total 
amplitude. This requires a calculation of the (long-distance) imaginary part 
at least at the two-loop level which will introduce a scale dependence. 
In addition, the two-loop contribution is expected to be of the same order
of magnitude as the one-loop contribution which only appears at the level of 
the finite terms, as it will bring in a quadratically divergent term. 
This situation is analogous to the non-suppression of the one-loop 
contribution to the real part ($\sim \Delta$) with respect to the tree 
level. The two-loop contribution to the real part, on the other hand, is 
expected to be suppressed by at least a factor of $\delta$ with respect to 
the tree level and the one-loop contribution. This is analogous to the 
one-loop logarithmic and finite terms which are suppressed by a factor of 
$\delta$ with respect to the tree level. For the numerical analysis we will 
therefore consider only the real part of the matrix elements [see the second 
expression in Eq.~(\ref{amp2})] using the experimental values of the final 
state interaction phases, $\delta_0^{\mbox{\tiny exp}}=(37\pm 3)^\circ$ 
and $\delta_2^{\mbox{\tiny exp}}=(-7\pm 1)^\circ$~\cite{phases}. This 
procedure has also been followed in Ref.~\cite{BEFL}. However, as the 
imaginary part is a loop effect (suppressed by a factor of $\delta$ with 
respect to the tree level contribution), its effect on the absolute value of 
the amplitude strictly speaking is of the two-loop order. Consequently, we 
will also compare our results with the ones obtained by taking the 
(long-distance) imaginary part to zero, i.e., by taking 
${\sum_i z_i\, \langle Q_i\rangle_I}={\sum_i z_i \,\mbox{Re}\langle 
Q_i\rangle_I}$. This holds for an estimate of the size of higher order 
effects which is generally disregarded in the~literature.

In Table~\ref{tab8} we show the numerical values of the amplitudes for various 
values of the matching scale and fixed values of $\Lambda_{\mbox{\tiny QCD}}=
\Lambda^{(4)}_{\overline{\mbox{\tiny MS}}}$ and the strange quark mass 
$m_s$. The numerical analysis is done using the leading logarithmic, as 
well as, the next-to-leading logarithmic values of the Wilson coefficients
listed in the appendix. The NLO values are scheme dependent and are 
calculated within naive dimensional regularization (NDR) and in the 
't Hooft-Veltman scheme (HV), respectively.\footnote{We are very thankful 
to M.~Jamin for providing us with the numerical values of the Wilson 
coefficients used in this section.} The difference between the two NLO 
results at a given scale reveals the uncertainty due to the lack of any 
reference to the renormalization scheme dependence in the effective 
low-energy calculation. 
\begin{table}[tbh]
\begin{eqnarray*}
\begin{array}{|c||c|c|c||c|c|c|}
\hline
& \multicolumn{3}{|c||}{\mbox{Re}a_0\,} & 
\multicolumn{3}{|c|}{\mbox{Re}a_2\,} \\ \cline{2-7}
\Lambda_c & \hspace*{5mm}\mbox{LO}\hspace*{5mm} & 
\hspace*{5mm}\mbox{NDR}\hspace*{5mm} & \hspace*{5mm}\mbox{HV}\hspace*{5mm} 
& \hspace*{5mm}\mbox{LO}\hspace*{5mm} & \hspace*{5mm}\mbox{NDR}\hspace*{5mm} 
& \hspace*{5mm}\mbox{HV}\hspace*{5mm} \\ 
\hline\hline
\rule{0cm}{5.5mm}
0.5\,\mbox{GeV} &  3.90 & 0.74 & 4.48 & 0.063  & 0.086  & 0.063 \\[0.5mm]
0.6\,\mbox{GeV} &  3.50 & 2.58 & 3.57 & 0.027  & 0.032  & 0.028 \\[0.5mm]
0.7\,\mbox{GeV} &  3.53 & 2.89 & 3.45 & -0.025 & -0.028 & -0.025 \\[0.5mm]
0.8\,\mbox{GeV} &  3.75 & 3.13 & 3.58 & -0.090 & -0.101 & -0.095 \\[0.5mm]
0.9\,\mbox{GeV} &  4.08 & 3.42 & 3.83 & -0.167 & -0.188 & -0.178 \\[0.5mm]
1.0\,\mbox{GeV} &  4.49 & 3.76 & 4.17 & -0.257 & -0.289 & -0.274 \\[0.7mm] 
\hline\hline
\mbox{exp.} & \multicolumn{3}{|c||}{3.33} & 
\multicolumn{3}{|c|}{0.15} \\
\hline
\end{array}
\end{eqnarray*}
\caption{$\mbox{Re}a_0$ and $\mbox{Re}a_2$ (in units of $10^{-4}\,\mbox{MeV}$)
for $m_s(1\,\mbox{GeV})=175\,$MeV, $\Lambda_{\mbox{\tiny QCD}}=
\Lambda^{(4)}_{\overline{\mbox{\tiny MS}}}=325\,$MeV, and various values 
of the matching scale $\mu=\Lambda_c$.\label{tab8}}
\end{table}
%
%
\noindent
\begin{figure}[t]
\centerline{\epsfig{file=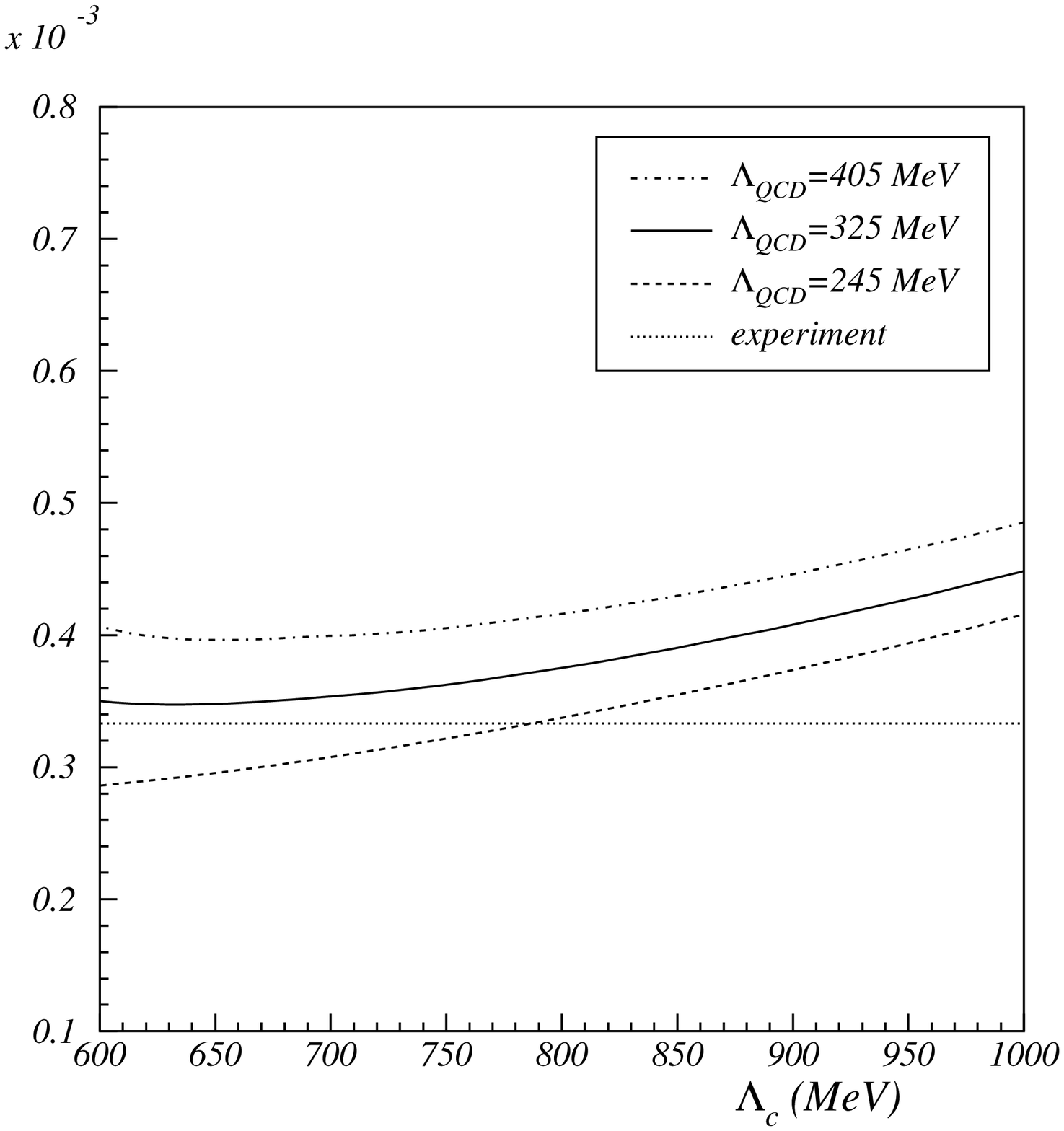,width=12cm}}
\vspace*{5mm}
\caption{$\mbox{Re}a_0$ (in units of MeV) with LO $z_i$ for $m_{s}(1\,
\mbox{GeV}) = 175\,$MeV and various values of $\Lambda_{\mbox{\tiny QCD}}$ 
as a function of the matching scale $\lc = \mu$.\label{knpi0}}
\end{figure}
\noindent
\begin{figure}[tb]
\vspace*{-2.0cm}
\centerline{\epsfig{file=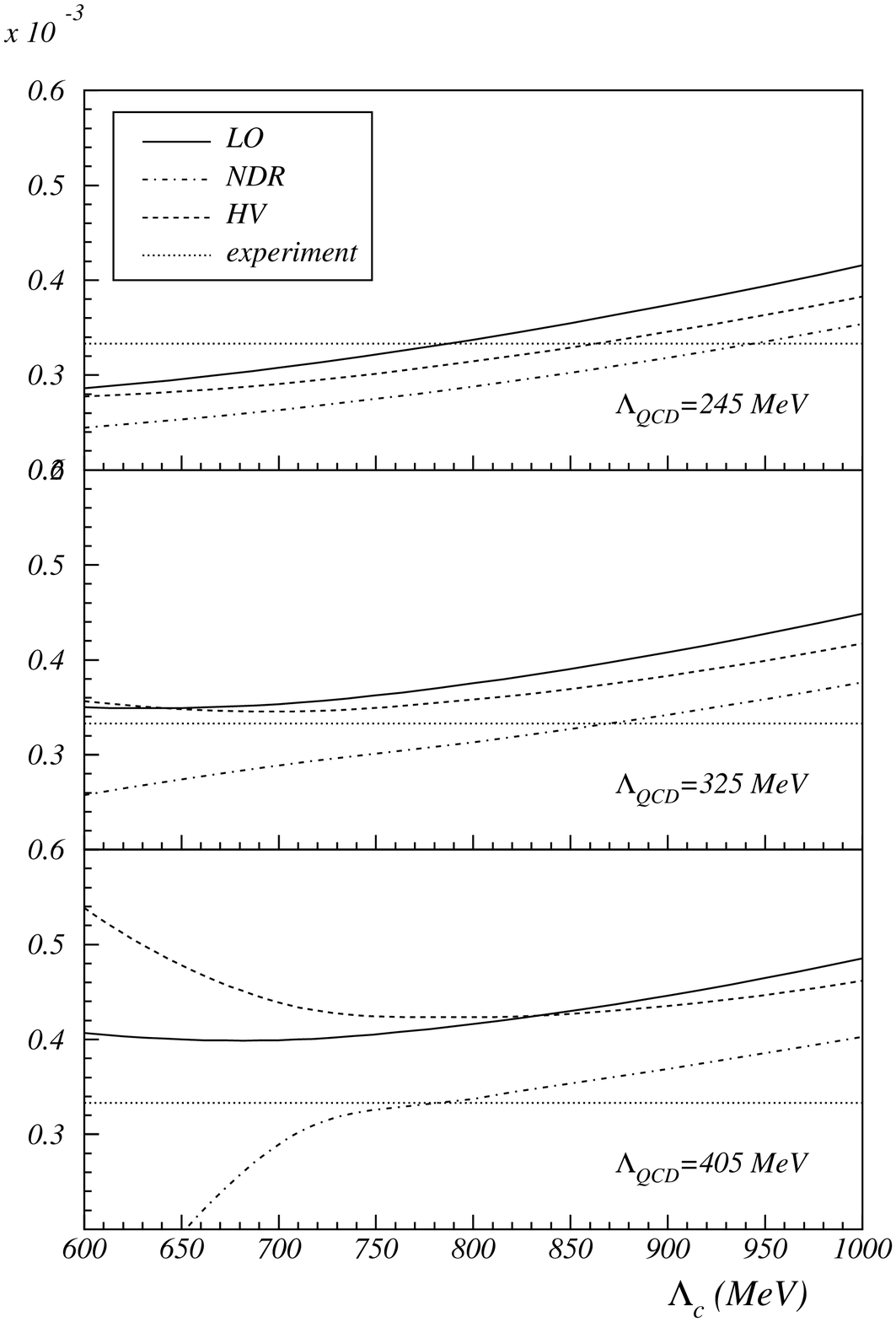,width=10.5cm}}
\vspace*{5mm}
\caption{$\mbox{Re}a_{0}$ (in units of MeV) with LO and NLO $z_i$ for 
$m_{s}(1\,\mbox{GeV}) = 175\,$MeV and various values of $\Lambda_{
\mbox{\tiny QCD}}$ as a function of the matching scale $\lc = \mu$.
\label{knpi0nlo}}
\end{figure}
\noindent
\begin{figure}[tb]
\centerline{\epsfig{file=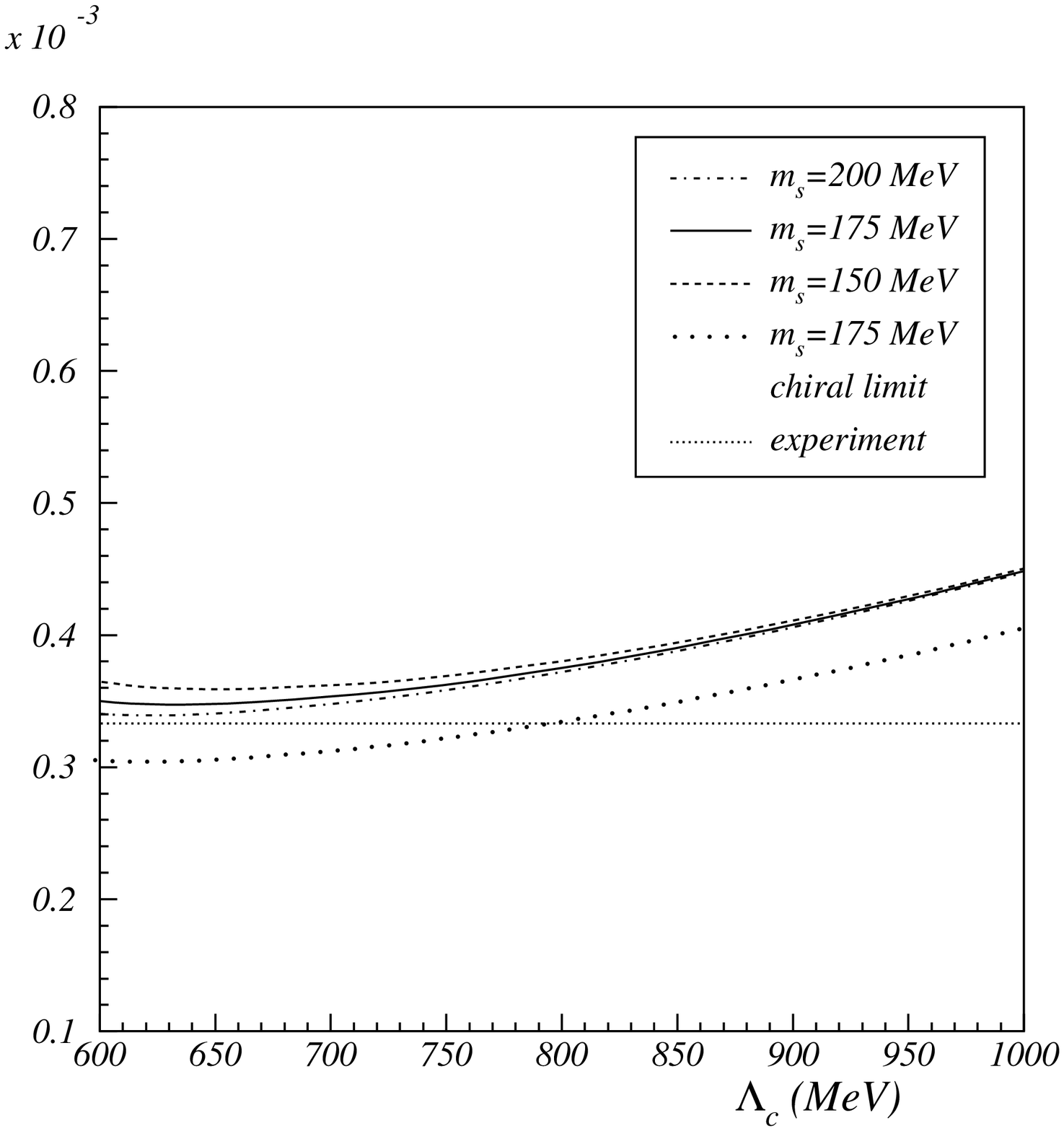,width=12cm}}
\vspace*{5mm}
\caption{$\mbox{Re}a_0$ (in units of MeV) with LO $z_i$ for 
$\Lambda_{\mbox{\tiny QCD}}=325\,$MeV and various values of $m_s(1\,
\mbox{GeV})$ as a function of the matching scale $\lc = \mu$.
\label{knpi0ms}}
\end{figure}

In Fig.~\ref{knpi0} we show $\mbox{Re}a_0$ calculated with leading order 
Wilson coefficients for various values of $\Lambda_{\mbox{\tiny QCD}}$ as 
a function of the matching scale. We take the (conservative) range of 
$\Lambda_{\mbox{\tiny QCD}}=325\pm 80\,\mbox{MeV}$ which corresponds to 
$\alpha_s(M_Z)=0.118 \pm 0.005$ \cite{AJB98}. First, we note that our 
result for $a_0$ shows an additional enhancement (around $30\,$-$\,50\,\%$ 
of the experimental value) compared to the result of Ref.~\cite{BBG3} which 
renders the amplitude in good agreement with the observed value for low 
values of the scale or even larger than the experimental value for large 
values of the scale. A significant enhancement arises from the $Q_1$ and 
$Q_2$ operators due to the modified matching prescription in the 
non-factorizable sector we discussed above. Numerically, at a scale of
$\Lambda_c=800\,\mbox{MeV}$ the modified momentum routing accounts for 
approximately $20\,\%$ of the final number(s) presented in Fig.~\ref{knpi0}. 
Another enhancement with respect to Ref.~\cite{BBG3} originates from the 
correction of the real part by the experimental phase [see Eq.~(\ref{amp2})]. 
Neglecting completely the effect of the ($\pi-\pi$) phase shift would reduce 
our result by a factor of $\cos\delta_0\simeq 0.8$. The remainder is due 
to the choice of the physical value $F_\pi$ instead of $f$ in the 
next-to-leading order terms of the factorizable and non-factorizable 
corrections. Our result depends only moderately on the matching scale 
although the stability falls off for large values of the scale around 
$1\,$GeV. We observe a cancellation between the scale dependence 
of the short- and long-distance contributions, i.e., the operator evolution 
in the quark picture is continued with the same pattern in the meson picture. 
The main uncertainty displayed in Fig.~\ref{knpi0} originates from the 
dependence of the Wilson coefficients on~$\Lambda_{\mbox{\tiny QCD}}$.
The uncertainty increases for very low values of the scale reflecting 
the poor perturbative behaviour expected at those scales especially for 
the large value of $\Lambda_{\mbox{\tiny QCD}}=405\,$MeV. Within the
(conservative) range of $\Lambda_{\mbox{\tiny QCD}}=325\pm 80\,$MeV we
considered, the value $405\,$MeV leads to the most distinct deviation 
from the experimental result which, however, does not exceed approximately 
$20\,\%$ of the observed value in the range $600\,\mbox{MeV}\lesssim 
\Lambda_c\lesssim 800\,\mbox{MeV}$ where the minimum occurs and the 
dependence on the scale is weak.

In Fig.~\ref{knpi0nlo} we compare the results for $\mbox{Re}a_0$ we obtain 
using the LO and NLO Wilson coefficients, respectively. In the HV scheme,
for moderate values of $\Lambda_{\mbox{\tiny QCD}}$ introducing the NLO 
coefficients does not significantly affect the numerical values of the 
$\Delta I=1/2$ amplitude which is found to be only slightly suppressed
with respect to the LO result. The main effect of the NLO coefficients 
is that they further reduce the dependence on the matching scale. This
statement does not hold within the NDR scheme. In this scheme, for 
$\Lambda_{\mbox{\tiny QCD}}=245\,\mbox{MeV}$ the effect of the NLO 
coefficients is also moderate but noticeably increases for large values 
of $\Lambda_{\mbox{\tiny QCD}}$ leading to a distinct suppression of 
the LO result. For values of $\Lambda_{\mbox{\tiny QCD}}$ as large as
405$\,\mbox{MeV}$ both the HV and the NDR results rapidly diverge for 
low values of the matching scale ($\lesssim 700\,\mbox{MeV}$) indicating
the loss of perturbativity. Taking into account the fact that we do not 
incorporate the effects of higher resonances and cannot adopt too high
values of the scale, a choice of $\Lambda_c$ around $700\,$-$\,800\,$MeV 
seems to be most appropriate. For $\Lambda_{\mbox{\tiny QCD}}=325\,\mbox{MeV}$
($245\,\mbox{MeV}$) the effect of the NLO coefficients is less pronounced,
and scales as low as $600\,$-$\,650\,\mbox{MeV}$ ($500\,\mbox{MeV}$), where 
the LO minimum occurs, appear to be acceptable. Above these scales the 
deviation of the NLO results from the experiment does not exceed 
$20\,$-$\,25\,\%$ of the experimental value. Moreover, the difference 
between LO and NLO (HV and NDR) values is moderate, of the order of at 
most $20\,$-$\,25\,\%$ of the observed value.\footnote{The comparison of 
the LO and NLO coefficients should be used with caution as it partly 
originates from a change in the value of the QCD coupling for a chosen 
value of $\Lambda_{\overline{\mbox{\tiny MS}}}$ \cite{BBL}.} In all the 
cases the tendency for a large enhancement of the required size  
remains~present. 

In Fig.~\ref{knpi0ms} we show the weak dependence of $\mbox{Re}a_0$ (with 
LO Wilson coefficients) on the strange quark mass which arises from the 
matrix element of the gluon penguin operator [$\langle Q_6\rangle_0\propto 
1/m_s^{2}\,$]. We notice that the contribution from $Q_6$ to the $\Delta 
I=1/2$ amplitude for small values of the cutoff ($\sim 600\,$MeV) roughly 
varies between $10\,$-$\,20\,\%$ of the total value and significantly 
decreases for large values of $\Lambda_c$. This behaviour is also found 
when the NLO coefficients are used. The effect of the remaining (penguin) 
operators is very small (below $1\,\%$ of the total result except for 
$Q_4$ which contributes at the level of $-3\,\%$). For comparison, in 
Fig.~\ref{knpi0ms} we also show $\mbox{Re}a_0$ calculated in the chiral limit. 
We observe that the result obtained in the chiral limit, for reasons explained 
above, is rather close to the numerically exact one, that is to say, the 
logarithmic and the finite terms in the non-factorizable corrections to the 
matrix elements are minor important provided that the matching scale is taken 
sufficiently large ($\Lambda_c\gtrsim 500\,$-$\,600\,\mbox{MeV}$). Finally, we 
note that the presence of the $\eta_0$ does not affect the numerical values 
of the amplitudes (in the octet limit the numbers given in Table~\ref{tab8} 
change by less than $1\,\%\,$).

\begin{figure}[tb]
\centerline{\epsfig{file=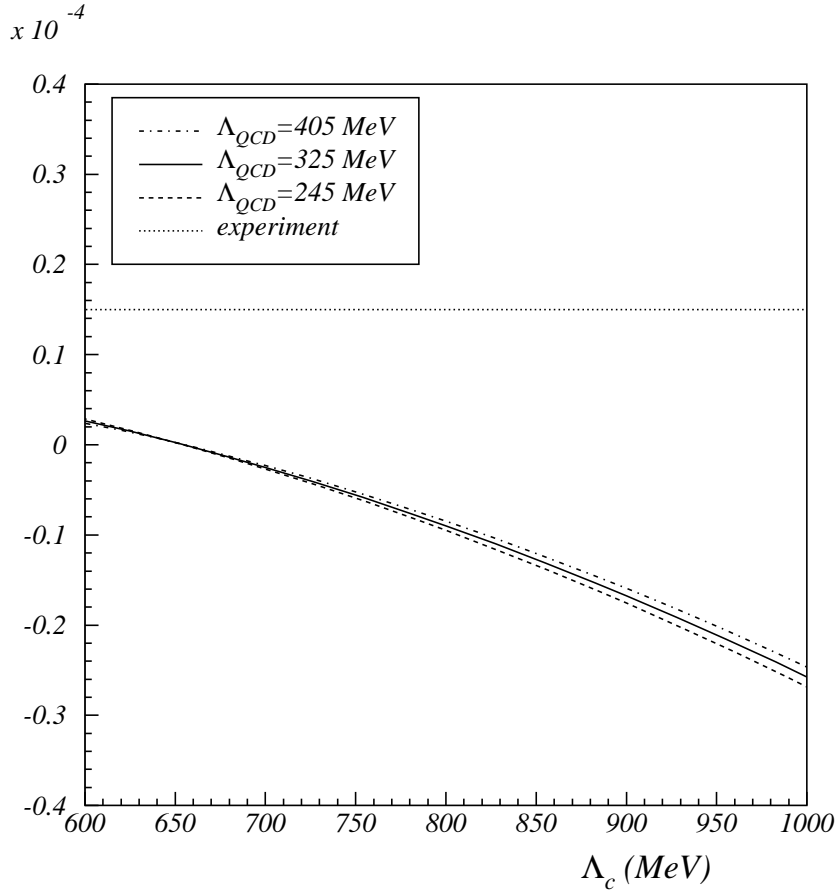,width=12cm}}
\vspace*{5mm}
\caption{$\mbox{Re}a_2$ (in units of MeV) with LO $z_i$ for various values
of $\Lambda_{\mbox{\tiny QCD}}$ as a function of the matching scale 
$\lc = \mu$.\label{knpi2}}
\end{figure}

\begin{figure}[tb]
\vspace*{-1.65cm}
\centerline{\epsfig{file=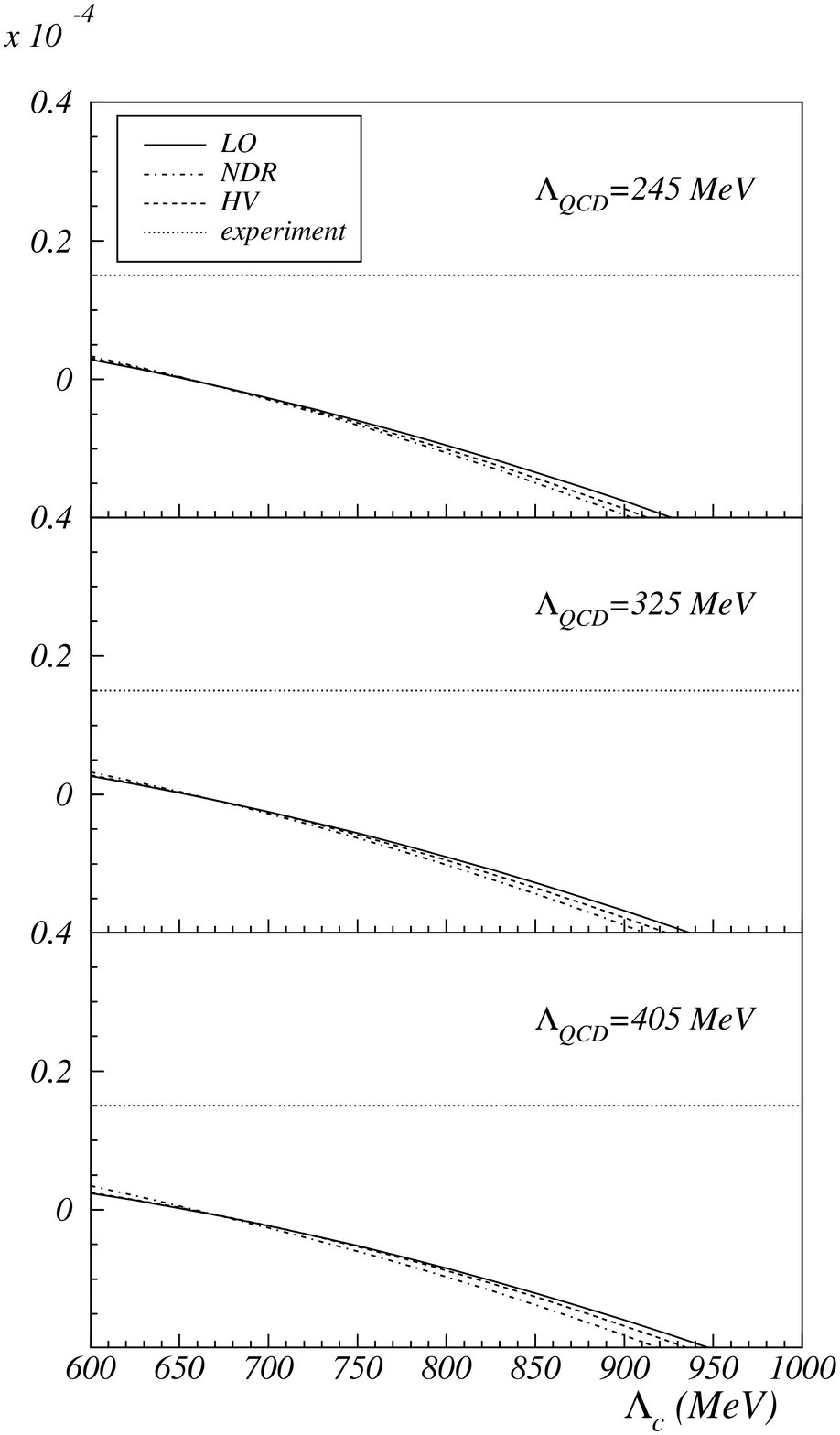,width=9.0cm}}
\vspace*{5mm}
\caption{$\mbox{Re}a_2$ (in units of MeV) with LO and NLO $z_i$ for various 
values of $\Lambda_{\mbox{\tiny QCD}}$ as a function of the matching scale 
$\lc = \mu$.\label{knpi2nlo}}
\end{figure}

In distinction to the $\Delta I=1/2$ amplitude, the $\Delta I=3/2$ 
amplitude depicted in Fig.~\ref{knpi2} (with LO Wilson coefficients) is 
highly unstable. In addition, the numerical values lie well below the 
measured value. The amplitude even changes sign [due to the large negative 
coefficient of the quadratic term in Eq.~(\ref{isq22})]. The large 
uncertainty can be understood, as we already discussed above, from the fact 
that the two numerically leading terms, the tree level and the one-loop 
quadratically divergent term, have approximately the same size but opposite 
sign. On the one hand, this property is generally welcomed as it explains the 
origin of the suppression of the $\Delta I=3/2$ amplitude which turns out 
to be sufficiently suppressed whatever the particular chosen scale is between 
$600\,$MeV and $900\,$MeV. On the other hand, the large cancellation implies 
that the result will be significantly affected by higher order terms which 
are expected to be of the order of the one-loop logarithmic and finite terms. 
We note that the agreement with the experimental value is not improved in 
the chiral limit. We also notice that the numerical values depicted in 
Fig.~\ref{knpi2} depend only weakly on the choice of $\Lambda_{\mbox{\tiny 
QCD}}$. In Fig.~\ref{knpi2nlo} we compare the results for $\mbox{Re}a_2$ we 
obtain using the LO and NLO Wilson coefficients, respectively. We observe 
that the effect of the NLO coefficients is negligible with respect to the 
large discrepancy between our results and the observed value. The small 
effect of the NLO coefficients indicates the validity of perturbation 
theory and further supports the supposition that the discrepancy is due 
the lack of accuracy in the low-energy part of the calculation.
\noindent
\noindent

\begin{figure}[tb]
\centerline{\epsfig{file=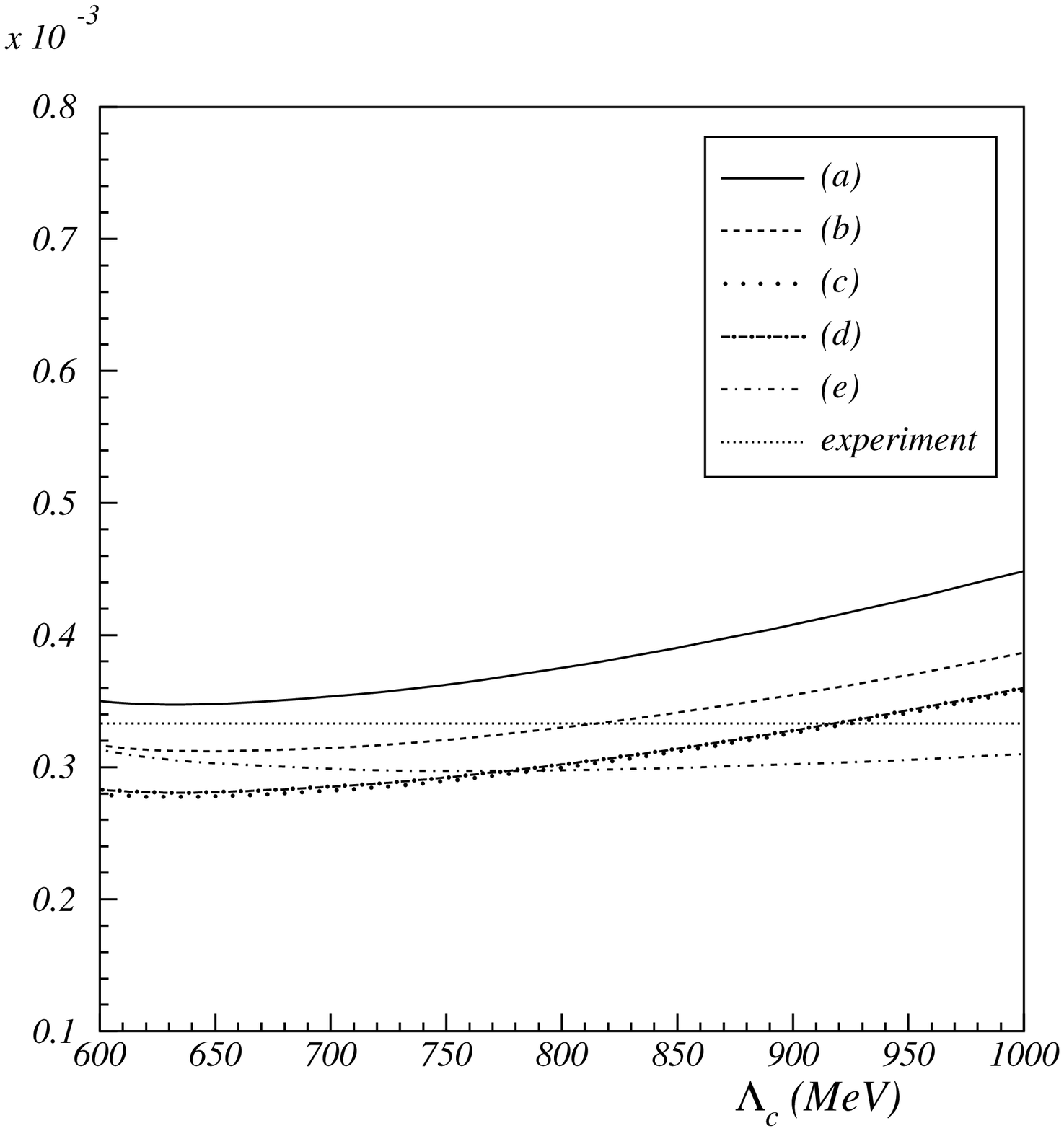,width=12cm}}
\vspace*{5mm}
\caption{$\mbox{Re}a_0$ (in units of MeV) with LO $z_i$ for 
$m_{s}(1\,\mbox{GeV}) = 175\,$MeV and $\Lambda_{\mbox{\tiny QCD}}=325
\,\mbox{MeV}$ within different treatments of higher order corrections 
as explained in the text.\label{k0com}}
\end{figure}

\begin{figure}[tb]
\centerline{\epsfig{file=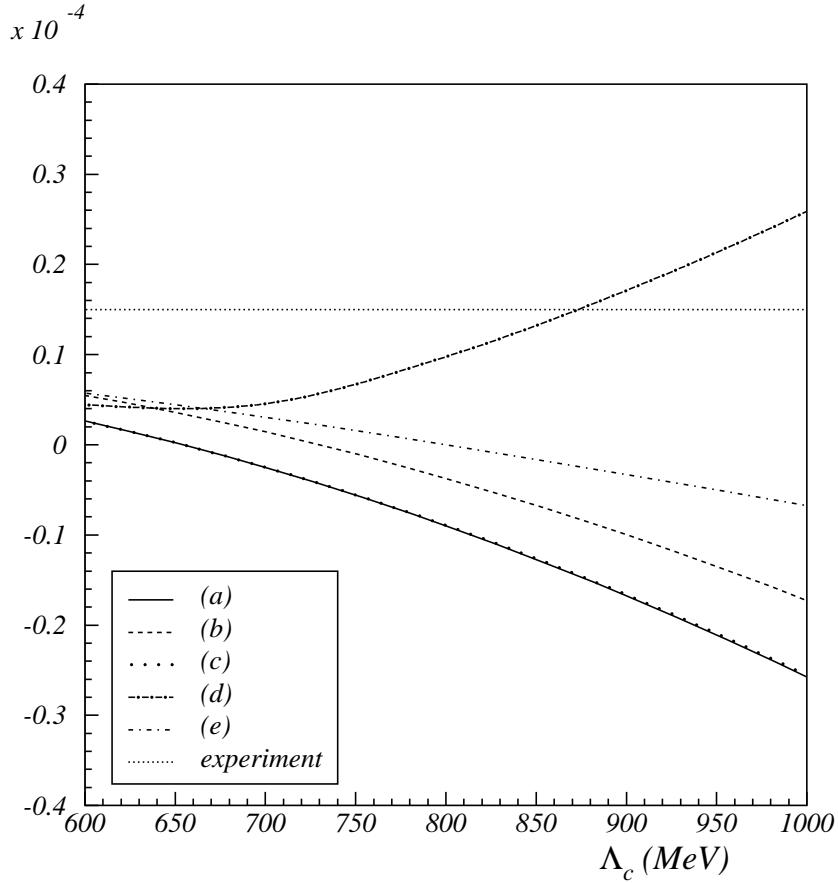,width=12cm}}
\vspace*{5mm}
\caption{$\mbox{Re}a_2$ (in units of MeV) with LO $z_i$ for 
$\Lambda_{\mbox{\tiny QCD}}=325\,\mbox{MeV}$ within different treatments 
of higher order corrections as explained in the text.\label{k2com}}
\end{figure}

The typical size of higher order effects in the calculation of the hadronic
matrix elements can be estimated in various ways. First, as we already
mentioned above, one may replace in all NLO terms the coefficient $1/F_\pi$ 
by $1/F_K$. The results obtained in this case [denoted by~(b)] are shown in 
Figs.~\ref{k0com} and \ref{k2com}. The $\Delta I=1/2$ amplitude is suppressed 
by approximately $20\,\%$ with respect to the result we obtained using 
$1/F_\pi$ [denoted by (a)] and is even in better agreement with the observed 
value. The $\Delta I=3/2$ amplitude, on the other hand, is enhanced but still 
far too much suppressed. Another estimation of higher order effects can be 
done, as we explained above, by completely neglecting the imaginary part of 
the matrix elements (c). This suppresses $\mbox{Re}a_0$ by a factor of 
$\cos\delta_0^{\mbox{\tiny exp}}\simeq 0.8$ but does not affect 
$\mbox{Re}a_2$. Similarly the absolute value of the amplitudes can be 
calculated by taking directly the imaginary part from Tables~\ref{tab3} 
and~\ref{tab4} without using the experimental phases (d). This procedure 
suppresses $\mbox{Re}a_0$ in the same way as in the previous case but largely 
re-stabilizes $\mbox{Re}a_2$, indicating that the results obtained for the 
$\Delta I=3/2$ amplitude (unlike those obtained for $\mbox{Re}a_0$) indeed 
can be significantly affected by higher orders corrections. It is unlikely, 
however, that higher order terms alone can account for the large discrepancy 
between our result and experiment, and effects from higher resonances are also
expected to be non-negligible for the small $\Delta I=3/2$ amplitude. 
Finally, the coefficient $1/F_{\pi}$ in the next-to-leading order terms can 
also be replaced by the bare coupling $1/f$ as it was done in 
Ref.~\cite{BBG3}. Even though this would introduce an unphysical dependence 
on the factorizable scale, formally the difference also concerns higher order 
effects.\footnote{The relation between $F_\pi$ and $f$ is given 
in Eq.~(62) of Ref.~\cite{HKPSB} and we obtain $f=105,\,112,\,120,$ $128,\,
136,\,145\,$MeV for $\Lambda_c=500,\,600,\,700,\,800,\,900,\,1000\,$MeV, 
respectively.} We observe that this choice (e) leads to a result for 
$\mbox{Re}a_0$ which is approximately scale independent. It also gives 
a more stable result for $\mbox{Re}a_2$ which, however, still is too 
much suppressed.
\noindent
\noindent

In summary, in all cases we discussed above the $\Delta I=1/2$ amplitude 
is obtained around the measured value with an uncertainty of less than 
25$\,\%$ or in most cases even less than 15$\,\%$.\footnote{The only 
exception  to this is the case where the large value of $\Lambda_{\mbox{
\tiny QCD}}=405\,\mbox{MeV}$ is taken at LO or NLO (HV scheme) using a 
matching scale as high as $\,\sim 1\,\mbox{GeV}$. In this (unfavourable) case 
the deviation from the observed value can be as large as $35\,$-$\,40\,\%$.} 
The result for Re$a_0$ is consequently solid and presumably could be 
significantly affected only by higher resonances. In view of the good 
agreement with the experiment we obtained at the pseudoscalar level their 
effect a priori is expected to be small. The $\Delta I=3/2$ amplitude, on 
the other hand, though showing the qualitatively correct behaviour of being 
suppressed with respect to the VSA result, emerges too much suppressed and 
is very unstable. However, higher order corrections to the matrix elements 
have been estimated large and could re-enhance it. In the same way higher 
resonances could easily enhance the result obtained at the pseudoscalar level.
Vector mesons can be incorporated in a straightforward (however lenghty) way, 
and it would be very interesting to investigate their effect in the present 
calculation. This also would allow more safely to choose higher values 
for the matching scale for which the short-distance contributions are 
more reliable.

We close this section by a brief review of several other attempts which
have been made to explain the $\Delta I = 1/2$ rule using different methods 
for the computation of the hadronic matrix elements. Interesting tendencies 
for an enhancement of the $\Delta I=1/2$ channel were found in particular 
in Ref.~\cite{PdR} by integrating out the quark fields in a gluonic 
background and in Ref.~\cite{JP} in the framework of QCD sum rules at 
the level of the inclusive two-point function. In Ref.~\cite{NS} 
quantitative results reproducing both the $\Delta I=1/2$ and 
$\Delta I=3/2$ channels were obtained adopting the point of view that 
in addition to $1/N_c$ effects due to one-loop corrections (similar to 
those of Fig.~\ref{ccnf}) diquark states play an important role. The 
results for the $\Delta I=1/2$ amplitude obtained in the present approach 
suggest that there are no large diquark effects not already taken into 
account in the $1/N_c$ corrections we calculated. The $\Delta I=1/2$ rule 
has also been investigated in the framework of chiral perturbation theory 
\cite{KMW} and the chiral quark model \cite{chqm}. At the present state of 
these methods the ratio $1/\omega=22.2$ cannot be predicted but is used to 
fit parameters of the models. Very recently the matrix elements relevant
for the $\Delta I=1/2$ rule were studied in lattice QCD with improved
statistics \cite{PK}. The authors used lowest-order chiral perturbation 
theory to relate the matrix elements $\langle\pi\pi|Q_i|K^0\rangle$ to 
$\langle\pi^+|Q_i|K^+\rangle$ and $\langle 0|Q_i|K^0\rangle$ calculated 
on the lattice. The ratio of the amplitudes computed in this way 
confirms the significant enhancement of the $\Delta I=1/2$ channel
although systematic uncertainties preclude a definite answer. Whereas 
the $\Delta I=1/2$ amplitude is obtained larger than the experimental 
value by approximately $40\,\%$ (quenched ensemble\footnote{Quantitative 
estimates of quenching effects on the coefficients of the chiral logarithms 
in the one-loop contributions to the $K\rightarrow\pi\pi$ amplitudes were 
presented in Refs.~\cite{GL,Pal}. In Ref.~\cite{GL} finite volume effects 
on the lattice were also investigated.}, $\beta=6.0$) the $\Delta I=3/2$ 
amplitude suffers from ambiguities in the choice of the meson mass due 
to the ignorance of higher order chiral corrections to the relation between 
$\mbox{Re}a_2$ and the $B_K$ parameter. Taking the meson mass $M^2=(m_K^2
+m_\pi^2)/2$ and using the quenched value of $B_K$ in the continuum limit 
the authors obtain a value for $\mbox{Re}a_2$ which also over-estimates 
the data by approximately $40\,\%$. The ratio of the amplitudes exhibits 
a strong dependence on the meson mass (see Fig.~11 of Ref.~\cite{PK}) due 
to the chiral behaviour of $\mbox{Re}a_2$. In lattice perturbation theory 
unlike in analytical methods, the matching of the renormalized operators to 
the Wilson coefficients can be rigorously done, at least in principle (see 
e.g. Ref.~\cite{Mart} and references therein). On the other hand, analytical 
methods like the $1/N_c$ approach followed in this paper allow for a direct 
evaluation of the $K\rightarrow\pi\pi$ amplitudes without the need of using 
reduction formulas to relate these amplitudes to the off-shell $K\rightarrow
\pi$ amplitudes (for this point see also Ref.~\cite{BPP} and references 
therein). 

While this paper was written an analysis of the $\Delta I=1/2$ rule 
was published \cite{BPdelta} which follows similar lines of thought as 
our work. In their analysis the authors used the $1/N_c$ expansion in the 
chiral limit in the framework of chiral perturbation theory and the Extended 
Nambu-Jona-Lasinio model, respectively. We agree on the coefficients of the 
quadratically divergent terms in the $1/N_c$ corrections to the matrix 
elements quoted therein. In the present analysis we did not investigate 
the method proposed in Ref.~\cite{BPdelta} to treat the scheme dependence 
appearing at the next-to-leading logarithmic order.
%
%
\section{$K^0 - \bar{K}^0$ Mixing}
The contributions from short-distance physics to $K^0 - \bar{K}^0$ 
mixing can be calculated from an effective $\Delta S = 2$ hamiltonian,
valid below the charm threshold, in which the heavy degrees of freedom 
are integrated out \cite{BJW},
\begin{equation}
{\cal H}_{ef\hspace{-0.5mm}f}^{\Delta S = 2}\,\,=\,\,{\cal F}(m_{t}^{2},
m_{c}^{2},M_{W}^{2},V_{\mbox{\tiny CKM}}) \,G_{F}
\left[\alpha_{s}(\mu)\right]^{-2/9} 
\left[1+\frac{\alpha_s(\mu)}{4\pi}J_3\right]\,{\cal O}_{\Delta S = 2}\,,
\label{heffss}
\end{equation}
where ${\cal O}_{\Delta S = 2}$ is the following four-quark operator:
\begin{equation}
{\cal O}_{\Delta S = 2}\,\,=\,\,\bar{s}_L\gamma^\mu d_L\,\,\bar{s}_L
\gamma_\mu d_L\,,
\end{equation}
with $\alpha_{s}(\mu)$ being the QCD running coupling with three active 
flavors and $J_3$ a renormalization scheme dependent coefficient appearing
at the next-to-leading logarithmic order.
${\cal F}(m_{t}^{2},m_{c}^{2},M_{W}^{2},V_{\mbox{\tiny CKM}})$ is a known 
function of the heavy quark masses, the $W$ boson mass, and CKM matrix 
elements. It incorporates the basic electroweak (box diagram) loop 
contributions \cite{IL}, as well as, the perturbative QCD effects described 
through the correction factors $\eta_1,\eta_2,\eta_3$ which have been 
calculated at the leading logarithmic \cite{GWGP,etlo} and the next-to-leading 
logarithmic order \cite{BJW,HN}. Terms depending on $\alpha_s(\mu)$
are factored out explicitly to exhibit the renormalization scale 
(and scheme) dependence of the coefficients which has to cancel the 
corresponding scale (and scheme) dependence of the hadronic matrix 
element of ${\cal O}_{\Delta S = 2}$ \cite{BBL}. The short-distance 
hamiltonian for $\Delta S=2$ transitions in Eq.~(\ref{heffss}) dominates 
the indirect CP violation in the neutral kaon system parameterized by 
$\varepsilon$. Contributions to $K^0 - \bar{K}^0$ mixing changing 
strangeness by two units through two $\Delta S=1$ transitions at long 
distances which are relevant for the $K_L-K_S$ mass difference \cite{BGK} 
are not considered in this article.

The hadronic matrix element of ${\cal O}_{\Delta S = 2}$ is usually 
parameterized in terms of the $B_{K}$ parameter which quantifies the 
deviation from the value obtained in the vacuum saturation approximation:
\begin{equation}
\langle\bar{K}^{0} | {\cal O}_{\Delta S = 2}(\mu) | K^{0} \rangle\,\, 
=\,\, B_{K}(\mu)\,
\langle\bar{K}^{0} | {\cal O}_{\Delta S = 2} | K^{0} 
\rangle_{\scriptsize\mbox{VSA}}\,,
\label{bkdef}
\end{equation}
where
\begin{equation}
\langle\bar{K}^{0} | {\cal O}_{\Delta S = 2} | K^{0}
\rangle_{\scriptsize\mbox{VSA}}\,\, = \,\,\frac{4}{3}\,F_{K}^{2} \mk^{2}\,.
\end{equation}
It is convenient to introduce the renormalization group invariant parameter
\cite{BBL,Bkpar}
\begin{equation}
\hat{B}_{K}\,\,=\,\,B_{K}(\mu)\,\left[\alpha_{s}(\mu)\right]^{-2/9} 
\left[1+\frac{\alpha_s(\mu)}{4\pi}J_3\right]\,,
\hspace*{1cm}
J_3\,=\,
\left\{ \begin{array}{ll}  \frac{307}{162}\,\, & (\mbox{NDR}) 
\\[2mm] \frac{91}{162} & (\mbox{HV}) \end{array} \right.\,, \label{bkhat}
\end{equation}
in which the scale (and scheme) dependences of the long- and short-distance 
contributions cancel within an exact realization of both perturbative and 
non-perturbative QCD. However, from the results for the $\Delta I=3/2$
$K\rightarrow\pi\pi$ amplitude discussed in the previous section we do 
not expect that the $\hat{B}_K$ we will obtain within the pseudoscalar
approximation used in the low-energy calculation will exhibit a negligible
dependence on the matching scale; the 27-plet operators which induce 
$\Delta S=1$ ($\Delta I=3/2$) and $\Delta S=2$ transitions are components 
of the same irreducible tensor under $SU(3)_L\times SU(3)_R$, that is to 
say, to leading order in the chiral expansion the $K^0-\bar{K}^0$ amplitude
can be related to the $\Delta I=3/2$ part of the $K\rightarrow\pi\pi$
amplitude using $SU(3)$ symmetry \cite{Don,jmg}. Consequently, we expect 
a similar pattern, i.e., a large negative quadratic term in the $1/N_c$ 
corrections to the matrix element which partly cancels the tree level 
contribution and renders the result more sensitive to corrections from higher 
order terms and higher resonances. On the other hand, we expect $SU(3)$ 
breaking effects in $\Delta S=2$ transitions to be more pronounced than 
in $\Delta S=1$ transitions \cite{BSW}. In the following we will see that 
the $1/N_c$ expansion restricted to the pseudoscalar mesons indeed leads 
to a significantly scale dependent result for $\hat{B}_K$. However, the
scale dependence is less pronounced than the one of the $\Delta I=3/2$
amplitude due to corrections beyond the chiral limit. Finally, as we 
already discussed above, the low-energy calculation does not allow any 
reference to the renormalization scheme dependence. Nevertheless, a 
comparison of the $\hat{B}_K$ parameter obtained from the LO and NLO 
coefficient function of ${\cal O}_{\Delta S=2}$ can be used to to test 
the validity of perturbation theory and to estimate the uncertainties 
arising from the short-distance part.
%
%
\subsection{Factorizable Loop Corrections}
\noindent
\begin{figure}[t]
\centerline{\epsfig{file=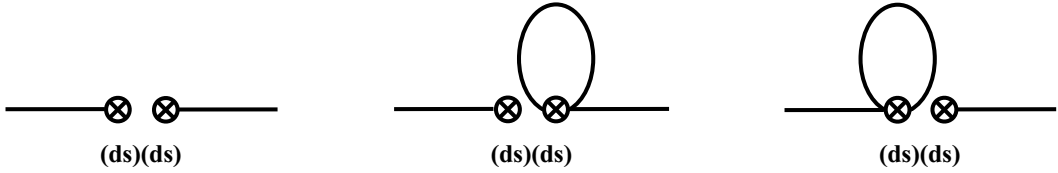,width=14cm}}
\vspace*{5mm}
\noindent
\caption{Factorizable contributions to the matrix element of the 
$K^0-\bar{K^0}$ mixing amplitude in the isospin limit.\label{bkf1}}
\end{figure}
%
%
To obtain the factorizable non-perturbative corrections to the $\Delta S = 2$ 
transition we have to calculate the diagrams in Fig.~\ref{bkf1}. Using the
chiral representation of the quark current in Eq.~(\ref{curr}) and reducing 
the result to the basic integrals listed in Appendix~B of Ref.~\cite{HKPSB}
we obtain the unrenormalized (bare) matrix element:
\begin{eqnarray}
\lefteqn{\langle \bar{K}^{0} | {\cal O}_{\Delta S = 2} | K^{0}
\rangle^{F}_{(0)}\,\,\, = \,\,\, \mk^2 f^2 \Bigg[ 1 + \frac{16 L_{5}}{f^2}
\mk^2} \nonumber \\
&& \hspace*{4mm} - {1\over{9 f^{2}}}\,\Big( \left( a + 2\,b \right)^{2} \,
I_{1}[\me] + 2\,\left( a - b \right)^{2} I_{1}[\mep] + 18\,I_{1}[\mk] 
+ 9\,I_{1}[\mp] \Big) \Bigg]\,,\hspace*{3mm} 
\label{bkfdintegral}
\end{eqnarray}
with $a$ and $b$ defined in Eq.~(\ref{abth}). Multiplying 
Eq.~(\ref{bkfdintegral}) with $Z_K^{-1}$, i.e., including a factor 
$Z_K^{-1/2}$ for each external kaon field (compare Eqs.~(16) and (59) of 
Ref.~\cite{HKPSB}), we arrive at
\begin{eqnarray}
\lefteqn{\langle \bar{K}^{0} | {\cal O}_{\Delta S = 2} | K^{0}
\rangle^{F}\,\,\,=\,\,\, 
\mk^2 f^2 \Bigg[ 1 + \frac{8 L_{5}}{f^{2}} \mk^2} \nonumber \\
&& \hspace*{4mm}
-\frac{1}{12 f^{2}} \Big( 9\,I_{1}[\mp] + 18\,I_{1}[\mk]
+\left( a + 2\,b \right)^{2} I_{1}[\me] + 2\,\left( a - b \right)^{2} 
I_{1}[\mep] \Big)\Bigg]\,. \hspace*{3mm} \label{fmeren}
\end{eqnarray}
Comparing Eq.~(\ref{fmeren}) with Eqs.~(26) and (63) of Ref.~\cite{HKPSB}
we observe that the correction factor in the brackets which is due to the 
higher order (factorizable) contributions to the matrix element  is 
completely absorbed (including the finite terms) in the renormalization of 
the kaon decay constant, as it is required by current conservation, leading 
to the final result for the (renormalized) factorizable matrix element
\begin{equation}
\langle \bar{K}^{0} | {\cal O}_{\Delta S = 2} | K^{0}\rangle^{F}_{(r)}
\,\,=\,\, \mk^2 F_K^2\,.\label{finfme}
\end{equation}
Eq.~(\ref{finfme}) represents the large-$N_c$ limit for the $K^0-\bar{K}^0$
matrix element, i.e., $B_K^{N_c\rightarrow\infty}=3/4$, to be compared with 
the VSA value one.
%
%
\subsection{Non-factorizable Loop Corrections}
\noindent
\begin{figure}[t]
\centerline{\epsfig{file=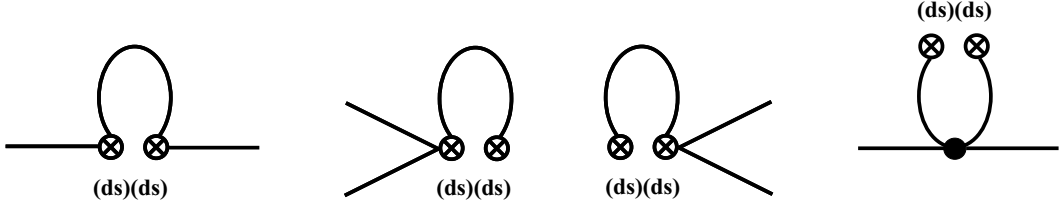,width=14cm}}
\vspace*{5mm}
\noindent
\caption{Non-factorizable contributions to the matrix element of the
$K^0-\bar{K^0}$ mixing amplitude in the isospin limit.\label{bknf1}}
\end{figure}
%
%
The $1/N_c$ corrections to Eq.~(\ref{finfme}) can be calculated from the 
non-factorizable loop diagrams depicted in Fig.~\ref{bknf1}. We determine 
the loop momenta along the lines developed in Section~2, that is to say, by
associating the cutoff to the effective color singlet boson connecting the
two currents. The simple structure of the non-factorizable diagrams makes
it possible to specify the complete analytic result for the matrix element 
in terms of loop integrals. In the $SU(2)$ limit the expression in which 
the integrals are reduced to the basic ones reads
\begin{eqnarray}
\lefteqn{\langle \bar{K}^{0} | {\cal O}_{\Delta S = 2} | K^{0} 
\rangle^{\NF}\,\,=\,\,
\frac{\Lambda_c^4}{32\pi^2}\,+\,\frac{1}{6}\Big(4m_K^2-2p_K^2
-(\chi_2+\chi_3)\Big)I_1[m_K]} \nonumber \\
&&\hspace*{4mm}
\,-\,\frac{1}{6}\big(\chi_2+\chi_3+2m_K^2+2p_K^2\big)m_K^2
I_3[m_K,m_K,0]\,-\,\frac{1}{2}(p_K^2+m_\pi^2)I_2[m_\pi,p_K] \nonumber \\
&&\hspace*{4mm}
\,-\,\frac{3}{2}\cos^2\theta\,(p_k^2+m_\eta^2)I_2[m_\eta,p_K]\,
-\,\frac{3}{2}\sin^2\theta\,(p_k^2+m_{\eta'}^2)\,I_2[m_{\eta'},p_K]
\nonumber \\
&&\hspace*{4mm}
+\,\frac{1}{4}I_4[m_\pi,p_K]\,+\,\frac{3}{4}\cos^2\theta\, I_4[m_\eta,p_K]
\,+\,\frac{3}{4}\sin^2\theta\, I_4[m_{\eta'},p_K]\,.
\label{bknfdintegral}
\end{eqnarray}
Here we replaced $a$ and $b$ by the $\eta-\eta'$ mixing angle $\theta$
and explicitly distinguished between the masses coming from the external 
kaon momentum, the explicit mass term in the lagrangian, and the propagators 
in the loops. In addition to the logarithmically and quadratically divergent 
integrals ($I_1,I_2,I_3$) listed in Appendix~B of Ref.~\cite{HKPSB} 
Eq.~(\ref{bknfdintegral}) contains the integral $I_4$ which exhibits 
a quartic dependence on the cutoff. Following the steps discussed in 
Ref.~\cite{HKPSB} we can give the analytic expression for $I_4$ in 
terms of a Taylor-series:
\begin{eqnarray}
I_4[m,p]&=&\D\lf\Integ\frac{q^2}{(q-p)^2-m^2} \nonumber\\[2mm]
&=&\frac{1}{16\pi^2}\Bigg\{-\frac{1}{2}\lc^4+m^2
\left[\lc^2-m^2\log\left(1+\frac{\lc^2}{m^2}\right)\right]
\nonumber\\[1mm]
&&
+\frac{p^2m^2}{(\lc+m^2)^2}\left[\frac{3}{2}\lc^4+\lc^2m^2-(\lc^2+m^2)^2
\log\left(1+\frac{\lc^2}{m^2}\right)\right]\nn[1mm]
&&
+\frac{p^4\lc^6}{6(\lc^2+m^2)^4}(\lc^2-2m^2)
+\frac{p^6\lc^6 m^2}{2(\lc^2+m^2)^6}\left(\lc^2-\frac{2}{3}m^2\right)
\Bigg\}+{\cal O}(p^8)\,. \label{I4} \hspace*{5mm}  
\end{eqnarray}
We note that the logarithmically divergent integral $I_3$ in 
Eq.~(\ref{bknfdintegral}) only appears with vanishing external momentum
and therefore can be largely simplified compared to the general expression
in Eq.~(75) of Ref.~\cite{HKPSB}. From Eq.~(\ref{bknfdintegral}) one can 
easily calculate the divergent terms. Taking the external momentum 
on-shell we obtain
\begin{equation}
\langle \bar{K}^{0} | {\cal O}_{\Delta S = 2} | K^{0} 
\rangle^{\NF}\,\,=\,\,  m_K^2F_K^2\left[ -\frac{3\lc^2}{(4\pi)^2F_K^2}
+ \frac{( 4 \mk^4 - 2 \mk^2 \mp^2 + \mp^4) }{(4\pi)^2F_K^2 m_K^2}
\,\log\lc^2 + \cdots \right] \,,\label{bkdiv}
\end{equation}
where the tree level result is factored out and the ellipses denote the 
finite terms we do not specify analytically. We observe that the quartic 
dependence on the cutoff is cancelled as required by chiral symmetry.

To illustrate the effect of the modified momentum routing we also recalculate 
the non-factorizable loop contributions in the approach used by Bardeen {\it 
et al.} \cite{BBG4} who associated the cutoff to the momentum of the virtual 
meson in the loop diagrams (see also the discussion in Ref.~\cite{HKPSB}):
\begin{eqnarray}
\lefteqn{\langle \bar{K}^{0} | {\cal O}_{\Delta S = 2} | K^{0} 
\rangle^{\NF}_{\mbox{\tiny BBG}}\,\,=\,\,
-\frac{1}{12}\,\Bigg[ 2 \left( \chi_{2} + \chi_{3} - 2 \mk^2 \right) 
I_{1}[\mk]} \nonumber \\[1mm]
&& \hspace*{4mm}
+ 3 \left( \mk^2 + \mp^2 \right) I_{1}[\mp]  
+ 9\cos^2\theta\,\left( \mk^2 + \me^2 \right) I_{1}[\me] 
+ 9\sin^2\theta \nonumber \\
&& \hspace*{4mm}
\times\left( \mk^2 + \mep^2 \right) I_{1}[\mep] 
+ 2 \mk^2 \left( \chi_{2} + \chi_{3} + 4 \mk^2 \right) I_{3}[\mk,\mk,0]\, 
\Bigg]\,,\hspace*{4mm} \label{bkbbg}
\end{eqnarray}
where the external momentum is already taken on-shell. For comparison with
Eq.~(\ref{bknfdintegral}) in Eq.~(\ref{bkbbg}) we included the small effect
of the singlet $\eta_0$. Solving the integrals we obtain the divergent part 
of the non-factorizable loop corrections:
\begin{equation}
\hspace*{-0.8mm}\langle \bar{K}^{0} | {\cal O}_{\Delta S = 2} | K^{0} 
\rangle^{\NF}_{\mbox{\tiny BBG}}\,\,=\,\, m_K^2F_K^2
\left[ -\frac{2\lc^2}{(4\pi)^2F_K^2}+\frac{(4\mk^4-2\mk^2\mp^2+\mp^4)}
{(4\pi)^2F_K^2 m_K^2}\,\log\lc^2 + \cdots \right] \,,\label{bkolddiv}
\end{equation}
to be compared with Eq.~(\ref{bkdiv}). We note that the results obtained 
in both calculations differ with respect to the quadratic cutoff dependence, 
as well as, with respect to the finite terms we do not give explicitly here 
for brevity. 
%
%
\subsection{Numerical Results \label{kknum}}
\begin{figure}[t]
\centerline{\epsfig{file=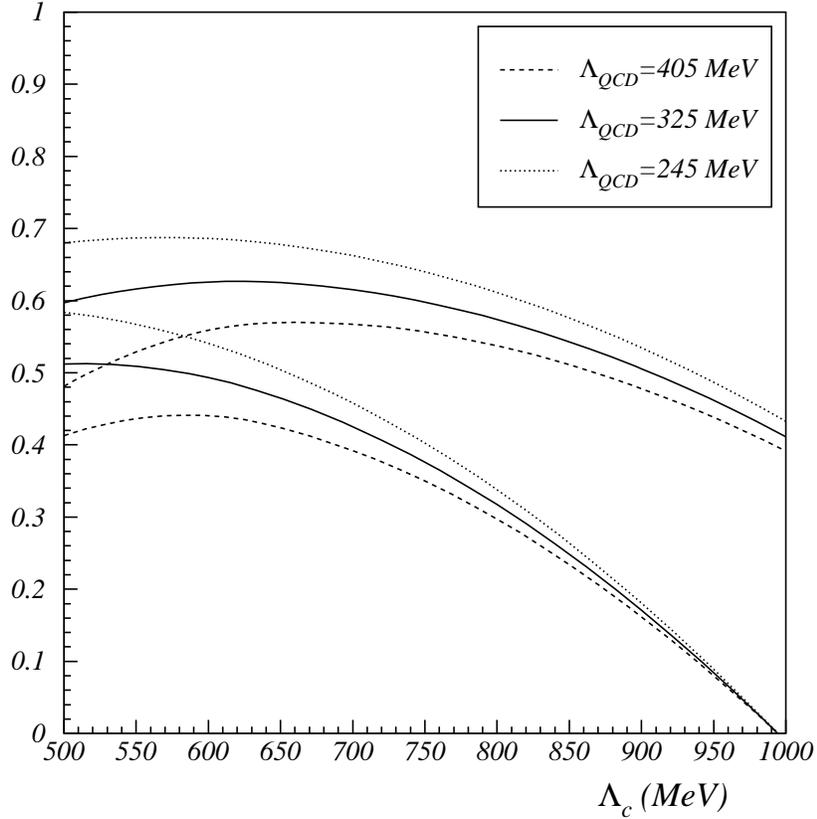,width=12cm}}
\vspace*{5mm}
\caption{$\hat{B}_K$ with LO Wilson coefficient for various values of 
$\Lambda_{\mbox{\tiny QCD}}$ as a function of the matching scale 
$\lc = \mu$. The lower set of curves shows the results of the present 
analysis, the upper set allows a comparison with Ref.~\cite{BBG4}. 
\label{kk1}}
\end{figure}

As a numerical input we use the values listed in Section~\ref{hme}. In 
Table~\ref{bknew} we show our results for the $K^0-\bar{K}^0$ matrix 
element and $B_K(\Lambda_c)$ obtained in the full calculation, i.e., 
including the effect of the $\eta_0$ in Eq.~(\ref{bknfdintegral}). In 
Fig.~\ref{kk1} we depict the renormalization group invariant parameter
$\hat{B}_K$ calculated with the leading order Wilson coefficient.
\begin{table}[tb]
\begin{eqnarray}
\begin{array}{|l||c|c|c|c|c|c|}\hline
\lc&0.5\,\mbox{GeV}&0.6\,\mbox{GeV}&0.7\,\mbox{GeV}&0.8\,\mbox{GeV}&0.9\,
\mbox{GeV}&1.0\,\mbox{GeV} \\ 
\hline\hline 
\rule{0cm}{5.0mm}
\langle {\cal O}_{\Delta S = 2} \rangle_{\,\mbox{\footnotesize tree}} 
& 3.14 & 3.14 & 3.14 & 3.14 & 3.14 & 3.14 \\[1mm]
\langle {\cal O}_{\Delta S = 2} \rangle_{\,\mbox{\footnotesize $\Lambda_c^2$}} 
& -1.17 & -1.68 & -2.29 & -2.99 & -3.78 & -4.67 \\[1mm]
\langle {\cal O}_{\Delta S = 2} \rangle_{\,\mbox{\footnotesize log$\,+\,$fin}} 
& 0.57 & 0.76 & 0.96 & 1.15 & 1.32 & 1.49 \\[1mm]
\hline\hline
\langle {\cal O}_{\Delta S = 2} \rangle 
& 2.54  & 2.22 & 1.81 & 1.30 & 0.68 &  -0.04\\[1mm]
\hline\hline
B_{K}(\lc) & 0.61 & 0.53 & 0.43 & 0.31 & 0.16 & -0.01 \\[1mm]
\hline
\end{array} \nonumber
\end{eqnarray}
\caption{Different contributions to the hadronic matrix element of
${\cal O}_{\Delta S = 2}$ (in units of $10^{9}\cdot\mbox{MeV}^{4}$)
and $B_{K}$, shown for various values of the cutoff $\Lambda_c$. 
\label{bknew}}
\end{table}
\noindent

The decrease of $B_{K}(\lc)$ with $\Lambda_c=\mu$ is qualitatively
consistent with the $\mu$ dependence of the coefficient function in
Eq.~(\ref{bkhat}), that is to say, the long-distance evolution counteracts
the evolution in the short-distance domain. This property is due to the
presence of the quadratic terms in the $1/N_c$ corrections which compensate
for the (weaker) increase of the logarithmic terms. However, the decrease
is found to be significant, and the scale dependence largely exceeds what 
is required to have an exact cancellation of both evolutions over a large 
range of the scale. As a result an acceptable stability of $\hat{B}_K$ is 
obtained only for low values of $\Lambda_c\simeq500\,$-$\,600\,\mbox{MeV}$.
The small values of $\hat{B}_K$ depicted in Fig.~\ref{kk1} (lower set of
curves) come from the negative coefficient of the quadratic term in 
Eq.~(\ref{bkdiv}) which is found to be enhanced by a factor of $3/2$ compared 
to the result of Ref.~\cite{BBG4}. This coefficient is the same as the one of 
the $\Delta I=3/2$ $K\rightarrow\pi\pi$ amplitude except for $SU(3)$ breaking 
effects (responsible for $F_K \neq F_\pi$) which reduce the negative slope 
of $\hat{B}_K$. As can be seen from Table~\ref{bknew}, the difference between 
the exact result and the one obtained in the chiral limit (i.e.,~in the absence
of chiral logarithms and finite terms) is more pronounced than in the case of 
the $K\rightarrow\pi\pi$ amplitudes. This is due mainly to the numerical
coefficient of the leading term ($\sim m_K^4$) in front of the logarithm in 
Eq.~(\ref{bkdiv}) which as expected is found larger in $\Delta S=2$ transitions
than in $\Delta S=1$ transitions. Because of the large positive coefficient 
the logarithmic term re-stabilizes $\hat{B}_K$ sizably with respect to the 
result obtained in the chiral limit. This also explains why the $\hat{B}_K$ 
parameter even if significantly scale dependent is much more stable than 
the $\Delta I = 3/2$ amplitude. The finite terms beyond the logarithms in 
Eq.~(\ref{bknfdintegral}) [i.e.,~beyond the~$\log(1+\lc^2/m^2)$ terms]
give a negative contribution to $B_K(\lc)$ roughly between $-0.05$ 
and $-0.08$ for $\lc$ around $600\,$-$900\,\mbox{MeV}$. Consequently, 
they are non-negligible in particular for large values of the scale where 
the cancellation between the tree level and the quadratic terms is large. 
Finally, we note that the presence of the $\eta_0$ does not significantly 
affect the numerical values of the $K^0-\bar{K}^0$ matrix element (in the 
octet limit the numbers given in Table~\ref{bknew} change by less than 
$3\,\%\,$).

To illustrate the effect of the momentum routing, in Fig.~\ref{kk1} we also 
show $\hat{B}_K$ obtained from Eq.~(\ref{bkbbg}) (upper set of curves). We 
use the same set of parameters as in Table~\ref{bknew} and also include the 
$\eta_0$. Comparing the two results we notice that $B_{K}(\lc)$ calculated 
within the modified momentum routing lies below the values found in the 
previous approach. Matching the long-distance results with the short-distance
contribution we observe that the $\hat{B}_K$ parameter obtained in the present 
analysis exhibits a significantly stronger dependence on the matching scale.
However, as we already discussed above, the quadratically divergent terms 
(and the finite terms) depend on the way we define the integration variable 
inside the loop. This can be seen from the different numerical factors in 
front of the quadratic terms in Eqs.~(\ref{bkdiv}) and~(\ref{bkolddiv}). 
Therefore we are forced to find a direct link between the short- and 
long-distance part of the calculation, as it is done by keeping track 
of the effective color singlet boson in both parts of the calculation. 
A consistent matching is then obtained by assigning the same momentum to 
the color singlet boson at long and short distances and by identifying 
this momentum with the loop integration variable (see Section~\ref{gen}).
This property is absent in the previous approach. The modification 
unambiguously determines the coefficient in front of the (quadratically 
and logarithmically) divergent terms and allows us to identify the 
ultraviolet cutoff of the long-distance terms with the short-distance 
renormalization scale $\mu$. Therefore we advocate the use of the 
modified matching prescription, even though the stability of our result 
is rather poor. The satisfactory stability obtained in Ref.~\cite{BBG4}
on the other hand is somehow inconclusive, as there is no underlying 
argumentation determining the quadratic terms. Our result also implies 
that the uncertainties due to the idealized identification of the cutoff 
$\lc$ with the upper limit of the meson momentum in the loop in 
Ref.~\cite{BBG4} might have been underestimated. In a complete meson 
theory the dependence on the momentum routing should be absent. However, 
as long as we are working in an effective low-energy approach as chiral 
perturbation theory we have to pay attention to this point.

Numerically, we find a range of acceptable stability in the energy 
regime from $500\,$MeV to $700\,$MeV (see Fig.~\ref{kk1}) leading to 
values for $\hat{B}_K$ in the range of $0.4 < \hat{B}_{K} < 0.6$.
The lower bound corresponds to a value of $\Lambda_{\mbox{\tiny QCD}}
= 405\,$MeV, whereas the upper bound corresponds to $\Lambda_{\mbox{\tiny 
QCD}} = 245\,$MeV. Comparing our result with the one of Ref.~\cite{BBG4} 
we observe a tendency for $\hat{B}_K$ to be decreased to values below $0.6$. 
This behaviour is due to the enhancement of the negative coefficient in front
of the quadratic term in the $1/N_c$ corrections to the $K^0-\bar{K}^0$ matrix
element and, to a smaller extend, also due to the finite terms omitted in 
Ref.~\cite{BBG4}. However, our result suffers from a sizable dependence on 
the matching scale which precludes a precise answer. 
\noindent
\begin{figure}[t]
\centerline{\epsfig{file=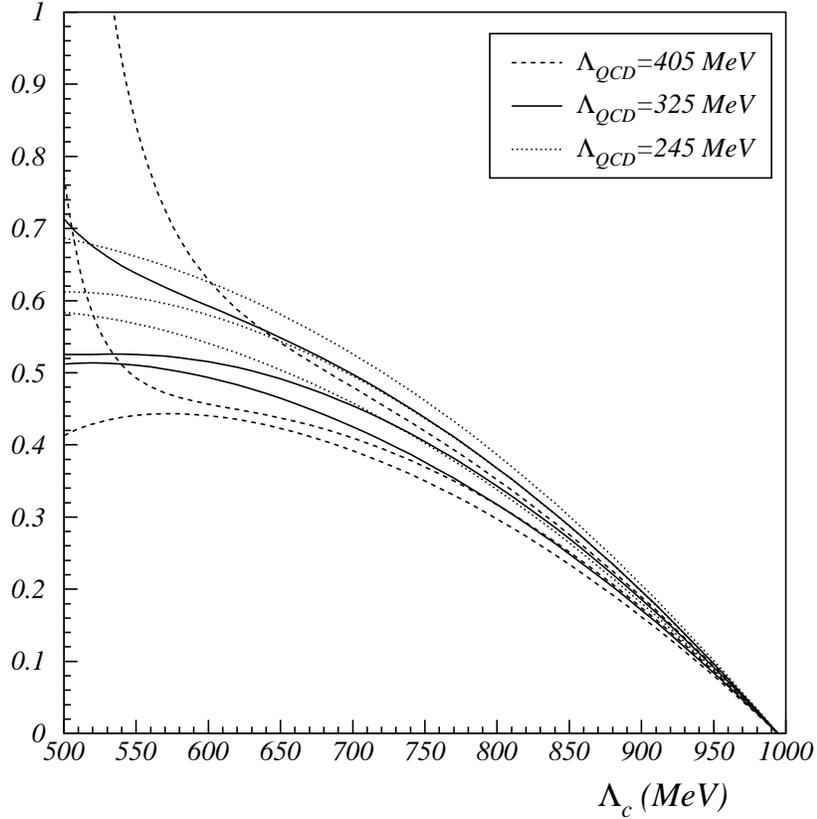,width=12cm}}
\vspace*{5mm}
\caption{$\hat{B}_K$ with LO and NLO Wilson coefficient for various values 
of $\Lambda_{\mbox{\tiny QCD}}=\Lambda^{(4)}_{\overline{\mbox{\tiny MS}}}$
as a function of the matching scale $\lc = \mu$. For each value of 
$\Lambda_{\mbox{\tiny QCD}}$ the lower (intermediate, upper) curve 
shows the LO (HV, NDR) result.\label{kk2}}
\end{figure}

In Fig.~\ref{kk2} we compare the results for $\hat{B}_K$ we obtain 
with the LO and NLO coefficient function. For $\Lambda_{\mbox{\tiny QCD}}
=325\,\mbox{MeV}$ in the HV scheme, introducing the NLO coefficient does 
not significantly affect the numerical values of the $\hat{B}_K$ parameter
which is found to be only slightly enhanced with respect to the LO result. 
In the NDR scheme, the effect of the NLO coefficient is also moderate 
for large values of the scale but noticeably increases for low values.
For very low values of $\lc\simeq 500\,\mbox{MeV}$ the NLO result can 
differ from the LO one by as much as 0.2. However, for these scales the 
scheme dependence increases rapidly and it is desirable to take (at least) 
a matching scale around $600\,$-$\,650\,\mbox{MeV}$ where $\hat{B}_K$ is 
still relatively smooth and roughly varies between $0.45$ and $0.6$. For 
$\Lambda_{\mbox{\tiny QCD}}=245\,\mbox{MeV}$ in both the HV and NDR schemes
a matching scale as low as $500\,\mbox{MeV}$ appears to be acceptable,
and within the range $\lc\simeq 500\,$-$\,650\,\mbox{MeV}$ $\hat{B}_K$ 
is obtained between $0.5$ and $0.7$. On the other hand we observe that 
the pseudoscalar approximation would simply fail if $\Lambda_{\mbox{\tiny 
QCD}}$ was found as large as $405\,\mbox{MeV}$, as a satisfactory 
perturbative behaviour is obtained only for $\lc\gtrsim 700\,\mbox{MeV}$, 
that is to say, for values of the scale where the stability of $\hat{B}_K$ 
is found to~be~poor. 

In summary, for values of $\Lambda_{\mbox{\tiny QCD}}\gtrsim 350\,\mbox{MeV}$
an estimate of $\hat{B}_K$ is hindered by the loss of perturbativity in the
range where the pseudoscalar approximation is expected to be valid, and for
lower values of $\Lambda_{\mbox{\tiny QCD}}$ (taking into account the scheme 
dependence) our calculation favours low values of $\hat{B}_K$ in the range 
\begin{equation}
0.4 \, < \, \hat{B}_{K} \, < \, 0.7\,.
\end{equation}
However, a satisfactory smooth behaviour is obtained only in a narrow 
range of the cutoff and, in addition, for values of the cutoff as low as 
the kaon mass or just above. Therefore the incorporation of higher resonances
is clearly required as for the $\Delta I = 3/2$ $K\rightarrow\pi\pi$ 
amplitude discussed above. On this issue, the analysis of the $\hat{B}_K$
parameter is similar to the one of the $\Delta I = 3/2$ amplitude, even if 
numerically the matching obtained for $\hat{B}_K$ is better than the one 
obtained for the $\Delta I = 3/2$ amplitude.

The $K^0-\bar{K^0}$ system has been studied in the past with various 
methods leading to different results for $\hat{B}_K$. The present status 
of quenched lattice calculations~[54\,-\,57] has been reviewed 
in Ref.~\cite{gupta}. The value reported by the author is $\hat{B}_K=0.86\pm 
0.06\pm 0.06$. Very recently the JLQCD Collaboration has presented a new 
analysis based on chiral Ward identities to non-perturbatively determine 
the mixing coefficients of the $\Delta S=2$ operator \cite{bklat}. The 
numerical results given in Ref.~\cite{bklat} are in agreement with the lattice
calculations quoted above. In the chiral quark model a value as high as 
$\hat{B}_K=1.1\pm 0.2$ has been obtained \cite{chqm}. Lower values for 
$\hat{B}_K$ have been found in the QCD hadronic duality approach \cite{bkdual}
($\hat{B}_K=0.39\pm 0.10$), by using $SU(3)$ symmetry and PCAC \cite{Don} 
($\simeq 1/3$), or using chiral perturbation theory at next-to-leading order 
\cite{bruno} ($0.42\pm 0.06$). QCD sum rules give results around 
$\hat{B}_K=0.5\,$-$\,0.6$ with errors in the range of $0.2\,$-$\,0.3$ 
\cite{bdg,bksum}. One might note that a value for $\hat{B}_K$ significantly 
below $0.7$ requires simultaneously high values of $|V_{ub}/V_{cb}|$ and 
$|V_{cb}|$ to be able to fit the experimental value of $\varepsilon$ 
\cite{BBL}. Finally, we note that the $\hat{B}_K$ parameter was also 
investigated in the framework of the $1/N_c$ expansion in Ref.~\cite{Bkpar}. 
In this work the matching was not performed at the level of the 
$K^0-\bar{K}^0$ matrix element but at the level of a related 2-point 
Green function. Numerically, the matching was found unsatisfactory good. 
We agree with this conclusion, as we discussed above, although in 
Ref.~\cite{Bkpar} the quadratic dependence on the UV cutoff was obtained 
in disagreement with the present analysis due to the use of a different 
momentum routing. This has been corrected very recently in 
Ref.~\cite{BPdelta}, and we agree with the results for the $1/N_c$ 
corrections to the $K^0-\bar{K}^0$ matrix element obtained there in the 
chiral limit. In the present paper we investigated the corrections beyond 
the chiral limit and found that they are sizable. On the other hand, the 
authors of Ref.~\cite{BPdelta} investigated higher order corrections 
calculated in the framework of the Extended Nambu-Jona-Lasinio model. 
As a result they obtained a better stability of the $\hat{B}_K$ parameter. 
This shows that corrections from higher order terms and higher resonances 
are expected to be large. Nevertheless the values of $\hat{B}_K$ we 
obtained in this analysis by performing a full calculation at the 
pseudoscalar level are meaningful and can be considered as reference 
values for further investigations incorporating the effects of higher 
resonances. 
%
%
\section{Conclusions}
The $1/N_c$ approach developed in Refs.~\cite{BBG3,BBG4} when modified 
along the lines of Ref.~\cite{HKPSB} leads to interesting results in the 
current-current sector of the $\Delta S=1$ and in the $\Delta S=2$
transitions. The main result of the present analysis is an additional 
enhancement of the $\Delta I =1/2$ channel in the $K\rightarrow\pi\pi$ 
amplitudes. This channel has been found sufficiently enhanced, in good 
agreement (with an accuracy of $80$ to approximately $100\,\%$) with 
the experiment, and widely stable over a large range of values of the 
matching scale roughly between $600\,$MeV and $900\,$MeV. 
It is certainly premature to say that the dynamical mechanism behind
the $\Delta I=1/2$ enhancement is completely understood. An agreement
at the level obtained in the present analysis a priori is not expected 
in an effective theory with only pseudoscalar mesons taken into account. 
Nevertheless we believe that the additional enhancement reported here is 
a further important indication that the $1/N_c$ approach can account for 
the bulk of the $\Delta I=1/2$ amplitude. This statement is also 
supported by the fact that higher order corrections both of short-distance 
origin and of long-distance origin at the pseudoscalar level, as we
discussed above, are not expected to largely affect the size of the 
$\Delta I=1/2$ enhancement. The agreement with the experiment also tends 
to show that the origin of the long-distance enhancement has to be found 
at the level of the pseudoscalar mesons and at energies below the rho
mass or even below the kaon mass. Certainly this has to be checked 
explicitly incorporating at least the effects of vector mesons.
We also believe that the $1/N_c$ approach can account for the bulk 
of the suppression of the $\Delta I=3/2$ channel. For this channel, 
however, the approximations made in the present analysis fell short 
of the desired accuracy. In particular, a large scale dependence has 
been found clearly requiring the incorporation of higher order terms 
and/or higher resonances. We note that the scale behaviour of the ratio 
of the two isospin amplitudes is dominated by the one of the 
$\Delta I = 3/2$ channel, and therefore it leads to a comparable 
uncertainty. Similarly, the $\hat{B}_K$ parameter suffers from a sizable 
dependence on the matching scale. Our calculation favours very low values 
of the scale ($\lesssim 700\,$MeV) leading to values for $\hat{B}_K$ in 
the range of $0.4<\hat{B}_K<0.7$. However, the large uncertainties associated 
with this result preclude a definite answer, and also make the incorporation 
of higher order terms and higher resonances very desirable.

\vspace{1cm}
\begin{center}{\large Acknowledgments}
\end{center}
We wish to thank Emmanuel Paschos for many valuable discussions. We also 
thank William Bardeen, Johan Bijnens, Jorge Fatelo, Jean-Marc G\'erard, 
and Gino Isidori for helpful comments. We are very thankful to Matthias 
Jamin for providing us with the numerical values of the Wilson coefficients 
used in Section~4 of this article. This work was supported in part by the 
Bundesministerium f\"ur Bildung, Wissenschaft, Forschung und Technologie 
(BMBF), 057D093P(7), Bonn, FRG, and DFG Antrag PA-10-1.

%
\newpage
\begin{appendix}
\section{Numerical Values of the Wilson Coefficients}
In this appendix we list the numerical values of the LO and NLO (HV and NDR)
Wilson coefficients for $\Delta S=1$ transitions used in Section~\ref{delta}. 
These values were supplied to us by M.~Jamin. Following the lines of 
Ref.~\cite{BJM} the coefficients $z_i$ are given for a 10-dimensional 
operator basis $\{Q_1,\ldots,Q_{10}\}$. Below the charm threshold the 
set of operators reduces to seven linearly independent operators [see
Eqs.~(\ref{qia})-(\ref{qio})] with
\begin{eqnarray}
Q_4\,=\,-Q_1+Q_2+Q_3\,,\hspace*{8mm}
Q_9\,=\,\frac{3}{2}Q_1-\frac{1}{2}Q_3\,,\hspace*{8mm} 
Q_{10}\,=\,\frac{1}{2}Q_1+Q_2-\frac{1}{2}Q_3\,. \label{linear-op} 
\end{eqnarray}
At next-to-leading logarithmic order in (renormalization group improved) 
perturbation theory the relations in Eq.~(\ref{linear-op}) receive 
${\cal O}(\alpha_s)$ and ${\cal O}(\alpha)$ corrections \cite{BJM,BBL}. 
In the present analysis we use the linear dependence at the level of 
the matrix elements $\langle Q_i\rangle_I$, i.e., at the level of the 
pseudoscalar representation where modifications to the relations in 
Eq.~(\ref{linear-op}) are absent. We note that the effect of the different 
treatment of the operator relations at next-to-leading logarithmic order 
which is due to the fact that in the long-distance part there is no 
(perturbative) counting in $\alpha_s$ is numerically negligible.

The following parameters are used for the calculation of the Wilson
coefficients:
\[
M_W\,=\,80.2\,\mbox{GeV},\hspace*{8mm}\sin^2\theta_W\,=\,0.23,\hspace*{8mm}
\alpha=1/129,
\]
\[
m_t\,=\,170\,\mbox{GeV},\hspace*{8mm}\overline{m}_b(m_b)\,=\,4.4\,\mbox{GeV},
\hspace*{8mm}\overline{m}_c(m_c)\,= \,1.3\,\mbox{GeV}\,.
\]
\vspace*{2mm}
\begin{eqnarray*}
\begin{array}{|c||c|c|c|c|c|c|}\hline
\mu&0.6\,\,\mbox{GeV}&0.7\,\,\mbox{GeV}&0.8\,\,\mbox{GeV}
&0.9\,\,\mbox{GeV}&1.0\,\,\mbox{GeV} \\ 
\hline\hline 
z_{1}  & -0.937 & -0.826 & -0.748 & -0.690 & -0.645 \\[0.2mm]
z_{2}  & 1.576 & 1.491 & 1.433 & 1.391 & 1.359 \\[0.2mm]
z_{3}  & 0.016 & 0.011 & 0.007 & 0.005 & 0.003 \\[0.2mm]
z_{4}  & -0.037 & -0.027 & -0.019 & -0.014 & -0.009 \\[0.2mm]
z_{5}  & 0.011 & 0.008 & 0.006 & 0.004 & 0.003 \\[0.2mm]
z_{6}  & -0.045 & -0.031 & -0.021 & -0.015 & -0.010 \\[0.2mm]
z_{7}/\alpha  & 0.023 & 0.017 & 0.012 & 0.008 & 0.005 \\[0.2mm]
z_{8}/\alpha  & 0.007 & 0.004 & 0.002 & 0.001 & 0.0004 \\[0.2mm]
z_{9}/\alpha  & 0.027 & 0.019 & 0.013 & 0.009 & 0.006 \\[0.2mm]
z_{10}/\alpha  & -0.006 & -0.003 & -0.002 & -0.001 & -0.0004 \\[0.2mm]
\hline
\end{array}
\end{eqnarray*}\\[2mm]\noindent
\centerline{Table~10: 
$\Delta S = 1$ LO Wilson coefficients for $\Lambda_{\mbox{\tiny QCD}}
=245\,\mbox{MeV}$.}
\newpage
\begin{eqnarray*}
\begin{array}{|c||c|c|c|c|c|c|}\hline
\mu&0.6\,\,\mbox{GeV}&0.7\,\,\mbox{GeV}&0.8\,\,\mbox{GeV}
&0.9\,\,\mbox{GeV}&1.0\,\,\mbox{GeV} \\ 
\hline\hline 
z_{1}  & -1.192 & -1.010 & -0.893 & -0.811 & -0.748 \\[0.2mm]
z_{2}  & 1.779 & 1.632 & 1.541 & 1.479 & 1.433 \\[0.2mm]
z_{3}  & 0.025 & 0.016 & 0.010 & 0.007 & 0.004 \\[0.2mm]
z_{4}  & -0.054 & -0.036 & -0.026 & -0.018 & -0.012 \\[0.2mm]
z_{5}  & 0.015 & 0.011 & 0.008 & 0.006 & 0.004 \\[0.2mm]
z_{6}  & -0.070 & -0.044 & -0.029 & -0.019 & -0.013 \\[0.2mm]
z_{7}/\alpha  & 0.033 & 0.023 & 0.017 & 0.012 & 0.008 \\[0.2mm]
z_{8}/\alpha  & 0.012 & 0.006 & 0.003 & 0.001 & 0.001 \\[0.2mm]
z_{9}/\alpha  & 0.040 & 0.027 & 0.019 & 0.013 & 0.008 \\[0.2mm]
z_{10}/\alpha  & -0.010 & -0.005 & -0.003 & -0.001 & -0.001 \\[0.2mm]
\hline
\end{array}
\end{eqnarray*}\\[2mm]\noindent
\centerline{Table~11:
$\Delta S = 1$ LO Wilson coefficients for $\Lambda_{\mbox{\tiny QCD}}
=325\,\mbox{MeV}$.}
\vspace*{1.5cm}
\begin{eqnarray*}
\begin{array}{|c||c|c|c|c|c|c|}\hline
\mu&0.6\,\,\mbox{GeV}&0.7\,\,\mbox{GeV}&0.8\,\,\mbox{GeV}
&0.9\,\,\mbox{GeV}&1.0\,\,\mbox{GeV} \\ 
\hline\hline 
z_{1}   & -1.576 & -1.246 & -1.065 & -0.947 & -0.861 \\[0.2mm]
z_{2}   & 2.104 & 1.824 & 1.676 & 1.582 & 1.517 \\[0.2mm]
z_{3}   & 0.041 & 0.023 & 0.014 & 0.009 & 0.006 \\[0.2mm]
z_{4}   & -0.082 & -0.051 & -0.034 & -0.023 & -0.015 \\[0.2mm]
z_{5}   & 0.022 & 0.015 & 0.010 & 0.007 & 0.005 \\[0.2mm]
z_{6}   & -0.119 & -0.066 & -0.041 & -0.026 & -0.016 \\[0.2mm]
z_{7}/\alpha   & 0.044 & 0.031 & 0.022 & 0.015 & 0.010 \\[0.2mm]
z_{8}/\alpha   & 0.024 & 0.010 & 0.005 & 0.002 & 0.001 \\[0.2mm]
z_{9}/\alpha   & 0.056 & 0.037 & 0.025 & 0.017 & 0.011 \\[0.2mm]
z_{10}/\alpha   & -0.017 & -0.008 & -0.004 & -0.002 & -0.001 \\[0.2mm]
\hline
\end{array}
\end{eqnarray*}\\[2mm]\noindent
\centerline{Table~12:
$\Delta S = 1$ LO Wilson coefficients for $\Lambda_{\mbox{\tiny QCD}}
=405\,\mbox{MeV}$.}
\newpage
\begin{eqnarray*}
\begin{array}{|c||c|c|c|c|c|c|}\hline
\mu&0.6\,\,\mbox{GeV}&0.7\,\,\mbox{GeV}&0.8\,\,\mbox{GeV}
&0.9\,\,\mbox{GeV}&1.0\,\,\mbox{GeV} \\ 
\hline\hline 
z_{1}   & -0.668 & -0.578 & -0.516 & -0.470 & -0.435 \\[0.2mm]
z_{2}   & 1.391 & 1.326 & 1.282 & 1.252 & 1.229 \\[0.2mm]
z_{3}   & 0.038 & 0.023 & 0.016 & 0.012 & 0.009 \\[0.2mm]
z_{4}   & -0.088 & -0.059 & -0.043 & -0.032 & -0.025 \\[0.2mm]
z_{5}   & 0.007 & 0.009 & 0.008 & 0.007 & 0.006 \\[0.2mm]
z_{6}   & -0.102 & -0.064 & -0.044 & -0.032 & -0.025 \\[0.2mm]
z_{7}/\alpha   & 0.018 & 0.012 & 0.008 & 0.006 & 0.005 \\[0.2mm]
z_{8}/\alpha   & 0.069 & 0.039 & 0.024 & 0.015 & 0.009 \\[0.2mm]
z_{9}/\alpha   & 0.045 & 0.029 & 0.020 & 0.014 & 0.010 \\[0.2mm]
z_{10}/\alpha   & -0.032 & -0.021 & -0.014 & -0.009 & -0.006 \\[0.2mm]
\hline
\end{array}
\end{eqnarray*}\\[2mm]\noindent
\centerline{Table~13:
$\Delta S = 1$ NLO Wilson coefficients (NDR) for 
$\Lambda_{\mbox{\tiny QCD}}=\Lambda^{(4)}_{\overline{\mbox{\tiny MS}}}
=245\,\mbox{MeV}$.}
\vspace*{1.5cm}
\begin{eqnarray*}
\begin{array}{|c||c|c|c|c|c|c|}\hline
\mu&0.6\,\,\mbox{GeV}&0.7\,\,\mbox{GeV}&0.8\,\,\mbox{GeV}
&0.9\,\,\mbox{GeV}&1.0\,\,\mbox{GeV} \\ 
\hline\hline 
z_{1}   & -0.898 & -0.739 & -0.644 & -0.579 & -0.531 \\[0.2mm]
z_{2}   & 1.569 & 1.444 & 1.373 & 1.326 & 1.292 \\[0.2mm]
z_{3}   & 0.033 & 0.019 & 0.012 & 0.007 & 0.005 \\[0.2mm]
z_{4}   & -0.060 & -0.038 & -0.025 & -0.017 & -0.011 \\[0.2mm]
z_{5}   & 0.012 & 0.008 & 0.006 & 0.004 & 0.003 \\[0.2mm]
z_{6}   & -0.060 & -0.036 & -0.024 & -0.016 & -0.010 \\[0.2mm]
z_{7}/\alpha   & -0.005 & -0.005 & -0.004 & -0.004 & -0.003 \\[0.2mm]
z_{8}/\alpha   & 0.046 & 0.027 & 0.017 & 0.011 & 0.007 \\[0.2mm]
z_{9}/\alpha   & 0.023 & 0.012 & 0.006 & 0.003 & 0.001 \\[0.2mm]
z_{10}/\alpha   & -0.038 & -0.024 & -0.016 & -0.010 & -0.007 \\[0.2mm]
\hline
\end{array}
\end{eqnarray*}\\[2mm]\noindent
\centerline{Table~14:
$\Delta S = 1$ NLO Wilson coefficients (HV) for $\Lambda_{\mbox{\tiny QCD}}
=\Lambda^{(4)}_{\overline{\mbox{\tiny MS}}}=245\,\mbox{MeV}$.}
\newpage
\begin{eqnarray*}
\begin{array}{|c||c|c|c|c|c|c|}\hline
\mu&0.6\,\,\mbox{GeV}&0.7\,\,\mbox{GeV}&0.8\,\,\mbox{GeV}
&0.9\,\,\mbox{GeV}&1.0\,\,\mbox{GeV} \\ 
\hline\hline 
z_{1}   & -0.805 & -0.712 & -0.623 & -0.558 & -0.509 \\[0.2mm]
z_{2}   & 1.495 & 1.424 & 1.359 & 1.312 & 1.278 \\[0.2mm]
z_{3}   & 0.095 & 0.046 & 0.027 & 0.018 & 0.013 \\[0.2mm]
z_{4}   & -0.193 & -0.104 & -0.068 & -0.048 & -0.035 \\[0.2mm]
z_{5}   & -0.019 & 0.005 & 0.009 & 0.009 & 0.008 \\[0.2mm]
z_{6}   & -0.261 & -0.121 & -0.072 & -0.049 & -0.035 \\[0.2mm]
z_{7}/\alpha   & 0.039 & 0.025 & 0.018 & 0.014 & 0.011 \\[0.2mm]
z_{8}/\alpha   & 0.181 & 0.079 & 0.042 & 0.024 & 0.014 \\[0.2mm]
z_{9}/\alpha   & 0.086 & 0.054 & 0.036 & 0.025 & 0.018 \\[0.2mm]
z_{10}/\alpha   & -0.056 & -0.034 & -0.021 & -0.013 & -0.008 \\[0.2mm]
\hline
\end{array}
\end{eqnarray*}\\[2mm]\noindent
\centerline{Table~15:
$\Delta S = 1$ NLO Wilson coefficients (NDR) for $\Lambda_{\mbox{\tiny QCD}}
=\Lambda^{(4)}_{\overline{\mbox{\tiny MS}}}=325\,\mbox{MeV}$.}
\vspace*{1.5cm}
\begin{eqnarray*}
\begin{array}{|c||c|c|c|c|c|c|}\hline
\mu&0.6\,\,\mbox{GeV}&0.7\,\,\mbox{GeV}&0.8\,\,\mbox{GeV}
&0.9\,\,\mbox{GeV}&1.0\,\,\mbox{GeV} \\ 
\hline\hline 
z_{1}   & -1.381 & -1.011 & -0.827 & -0.716 & -0.640 \\[0.2mm]
z_{2}   & 1.982 & 1.662 & 1.513 & 1.427 & 1.370 \\[0.2mm]
z_{3}   & 0.090 & 0.040 & 0.022 & 0.013 & 0.007 \\[0.2mm]
z_{4}   & -0.129 & -0.068 & -0.041 & -0.026 & -0.016 \\[0.2mm]
z_{5}   & 0.016 & 0.011 & 0.008 & 0.006 & 0.004 \\[0.2mm]
z_{6}   & -0.137 & -0.067 & -0.038 & -0.024 & -0.014 \\[0.2mm]
z_{7}/\alpha   & -0.008 & -0.003 & -0.002 & -0.002 & -0.002 \\[0.2mm]
z_{8}/\alpha   & 0.107 & 0.050 & 0.027 & 0.016 & 0.010 \\[0.2mm]
z_{9}/\alpha   & 0.052 & 0.027 & 0.015 & 0.009 & 0.005 \\[0.2mm]
z_{10}/\alpha   & -0.077 & -0.042 & -0.025 & -0.016 & -0.010 \\[0.2mm]
\hline
\end{array}
\end{eqnarray*}\\[2mm]\noindent
\centerline{Table~16:
$\Delta S = 1$ NLO Wilson coefficients (HV) for $\Lambda_{\mbox{\tiny QCD}}
=\Lambda^{(4)}_{\overline{\mbox{\tiny MS}}}=325\,\mbox{MeV}$.}
\newpage
\begin{eqnarray*}
\begin{array}{|c|c||c|c|c|c|c|c|}\hline
\mu&0.6\,\,\mbox{GeV}&0.7\,\,\mbox{GeV}&0.8\,\,\mbox{GeV}
&0.9\,\,\mbox{GeV}&1.0\,\,\mbox{GeV} \\ 
\hline\hline 
z_{1} & -0.176 & -0.795 & -0.738 & -0.657 & -0.592 \\[0.2mm]
z_{2} & 0.911 & 1.485 & 1.444 & 1.384 & 1.336 \\[0.2mm]
z_{3} & 0.350 & 0.108 & 0.052 & 0.030 & 0.019 \\[0.2mm]
z_{4} & -0.637 & -0.218 & -0.117 & -0.074 & -0.050 \\[0.2mm]
z_{5} & -0.318 & -0.027 & 0.004 & 0.009 & 0.009 \\[0.2mm]
z_{6} & -1.172 & -0.288 & -0.132 & -0.077 & -0.050 \\[0.2mm]
z_{7}/\alpha & 0.119 & 0.042 & 0.029 & 0.023 & 0.018 \\[0.2mm]
z_{8}/\alpha & 0.699 & 0.185 & 0.081 & 0.042 & 0.023 \\[0.2mm]
z_{9}/\alpha & 0.132 & 0.089 & 0.059 & 0.040 & 0.029 \\[0.2mm]
z_{10}/\alpha & -0.077 & -0.054 & -0.033 & -0.020 & -0.012 \\[0.2mm]
\hline
\end{array}
\end{eqnarray*}\\[2mm]\noindent
\centerline{Table~17:
$\Delta S = 1$ NLO Wilson coefficients (NDR) for $\Lambda_{\mbox{\tiny QCD}}
=\Lambda^{(4)}_{\overline{\mbox{\tiny MS}}}=405\,\mbox{MeV}$.}
\vspace*{1.5cm}
\begin{eqnarray*}
\begin{array}{|c||c|c|c|c|c|c|}\hline
\mu&0.6\,\,\mbox{GeV}&0.7\,\,\mbox{GeV}&0.8\,\,\mbox{GeV}
&0.9\,\,\mbox{GeV}&1.0\,\,\mbox{GeV} \\ 
\hline\hline 
z_{1}   & -2.603 & -1.494 & -1.102 & -0.901 & -0.778 \\[0.2mm]
z_{2}   & 3.138 & 2.084 & 1.739 & 1.573 & 1.475 \\[0.2mm]
z_{3}   & 0.370 & 0.102 & 0.044 & 0.023 & 0.012 \\[0.2mm]
z_{4}   & -0.403 & -0.140 & -0.072 & -0.042 & -0.025 \\[0.2mm]
z_{5}   & 0.035 & 0.014 & 0.010 & 0.007 & 0.005 \\[0.2mm]
z_{6}   & -0.463 & -0.141 & -0.067 & -0.037 & -0.021 \\[0.2mm]
z_{7}/\alpha   & -0.063 & -0.009 & -0.002 & -0.001 & -0.001 \\[0.2mm]
z_{8}/\alpha   & 0.342 & 0.105 & 0.048 & 0.026 & 0.014 \\[0.2mm]
z_{9}/\alpha   & 0.111 & 0.051 & 0.028 & 0.016 & 0.009 \\[0.2mm]
z_{10}/\alpha   & -0.179 & -0.078 & -0.042 & -0.024 & -0.014 \\[0.2mm]
\hline
\end{array}
\end{eqnarray*}\\[2mm]\noindent
\centerline{Table~18:
$\Delta S = 1$ NLO Wilson coefficients (HV) for $\Lambda_{\mbox{\tiny QCD}}
=\Lambda^{(4)}_{\overline{\mbox{\tiny MS}}}=405\,\mbox{MeV}$.}
\end{appendix}

\renewcommand{\textfraction}{1.0}
\newpage

\end{document}